\documentclass[aps,prl,twocolumn,superscriptaddress,nofootinbib,amsmath,amssymb,floatfix]{revtex4-2}

\usepackage{graphicx}
\usepackage{bm}
\usepackage{xcolor}
\usepackage{pifont}
\usepackage{hyperref}
\definecolor{prlblue}{HTML}{2E3092}
\hypersetup{colorlinks=true,allcolors=prlblue,
  pdftitle={Relaxation Critical Dynamics of Strong-to-Weak Spontaneous Symmetry Breaking},
  pdfauthor={Jiaqiang Li, Xianquan Yan, Han-Ze Li, Shuo Liu, Shuai Yin}}
\usepackage{cleveref}
\usepackage{algpseudocode}
\newcounter{algorithm}
\crefname{algorithm}{Algorithm}{Algorithms}
\Crefname{algorithm}{Algorithm}{Algorithms}
\crefname{equation}{Eq.}{Eqs.}
\Crefname{equation}{Equation}{Equations}
\creflabelformat{equation}{(#2#1#3)}
\crefrangelabelformat{equation}{(#3#1#4)--(#5#2#6)}

\crefname{figure}{Fig.}{Figs.}
\Crefname{figure}{Figure}{Figures}
\crefname{table}{Table}{Tables}
\Crefname{table}{Table}{Tables}
\crefname{section}{Sec.}{Secs.}
\Crefname{section}{Section}{Sections}
\crefname{subsection}{Sec.}{Secs.}
\Crefname{subsection}{Section}{Sections}

\definecolor{headcol}{HTML}{000000}
\newcommand{\runinhead}[1]{\textit{\textcolor{headcol}{#1.}}---}
\newcounter{mainsec}

\crefformat{mainsec}{#2\textit{#1}#3}
\Crefformat{mainsec}{#2\textit{#1}#3}
\newcommand{\mainsec}[2]{\par\renewcommand{\themainsec}{#2}\refstepcounter{mainsec}\label{#1}\runinhead{#2}}
\newcounter{emsec}
\renewcommand{\theemsec}{\Alph{emsec}}

\crefname{emsec}{End Matter}{End Matter}
\Crefname{emsec}{End Matter}{End Matter}
\newcommand{\emsec}[2]{\par\refstepcounter{emsec}\label{#1}\pdfbookmark[2]{\theemsec. #2}{emsec.\theemsec}\setcounter{equation}{0}\runinhead{\theemsec.\ #2}}

\makeatletter
\def\@bibstyle{apsrev4-2-arxiv}
\makeatother

\newcommand{\Tr}{\operatorname{Tr}}
\newcommand{\dket}[1]{|#1\rangle\!\rangle}

\newcommand{\Zt}{\mathbb{Z}_2}
\providecommand{\openone}{\mathbb{1}}
\newcommand{\SM}[1]{\cite[\cref{#1}]{SM}}
\newcommand{\Gs}{\mathcal G_\sigma}
\newcommand{\Gt}{\mathcal G_\tau}
\newcommand{\xini}{x_{\rm ini}}
\newcommand{\xobs}{x_{\rm obs}}

\begin{document}

\author{Jiaqiang Li}
\affiliation{Guangdong Provincial Key Laboratory of Magnetoelectric Physics and Devices, Sun Yat-sen University, Guangzhou 510275, China}
\affiliation{School of Physics, Sun Yat-sen University, Guangzhou 510275, China}%
\author{Xianquan Yan}
\affiliation{Department of Physics, National University of Singapore, Singapore 117551}
\affiliation{Department of Computer Science, National University of Singapore, Singapore 117417}

\author{Han-Ze Li}
\affiliation{Department of Physics, National University of Singapore, Singapore 117551}
\affiliation{Institute for Quantum Science and Technology, Shanghai University, Shanghai 200444, China}

\author{Shuai Yin}
\email{yinsh6@mail.sysu.edu.cn}
\affiliation{Guangdong Provincial Key Laboratory of Magnetoelectric Physics and Devices, Sun Yat-sen University, Guangzhou 510275, China}%
\affiliation{School of Physics, Sun Yat-sen University, Guangzhou 510275, China}%
\author{Shuo Liu}
\email{sl6097@princeton.edu}
\affiliation{Department of Physics, Princeton University, Princeton, NJ 08544, USA}
\title{Relaxation Critical Dynamics of
Strong-to-Weak Spontaneous Symmetry Breaking}

\begin{abstract}
Strong-to-weak spontaneous symmetry breaking (SW-SSB) is unique to mixed states: strong symmetry is broken while weak symmetry is preserved, requiring two order parameters.
We study the critical relaxation of both order parameters in a monitored $\Zt$ circuit, which maps exactly onto two nested bond-percolation problems. The initial state acts as a temporal boundary condition for each problem, wired if the state is ordered in the corresponding order parameter and free otherwise. 
An SW-SSB initial state wires only the strong-symmetry order parameter, a configuration that no pure state realizes.
Each autocorrelation exponent is the sum of an initial scaling dimension, set by the initial state, and an observation dimension, set by the observation time, whereas the equal-time correlator gives only the observation dimension. For the SW-SSB state, the initial dimension is $1/3$ for the weak-symmetry order parameter and $0$ for the strong-symmetry one, each at its own critical point. These initial dimensions also set how fast a stored bit is lost. The probability of decoding the initial value of a single qubit from the system alone decays as a power law, whereas system and environment together retain it. Stabilizer simulations confirm the theoretical prediction, and we propose an experimental protocol that measures these exponents without postselection.
Our results reveal a form of boundary criticality inaccessible to pure states.
\end{abstract}

\maketitle

\mainsec{sec:intro}{Introduction}An open system can realize a symmetry in two inequivalent ways~\cite{buca_note_2012,albert_symmetries_2014,lieu_symmetry_2020,lee_universality_2026,zhang_strong--weak_2025}.
A \emph{strong} symmetry $U$ acts on one side of the density matrix, $U\rho=e^{i\phi}\rho$, so that every state of the ensemble carries the same charge $e^{i\phi}$. A \emph{weak} symmetry acts on both sides, $U\rho U^\dagger=\rho$, and only the ensemble as a whole is invariant~\cite{de_groot_symmetry_2022,ma_average_2023,xue_tensor_2024,lessa_strong--weak_2025,wang_strong--weak_2026,ma_topological_2025}. A mixed state can spontaneously break the strong symmetry while preserving the weak one. This SW-SSB has no pure-state counterpart~\cite{wang_strong--weak_2026,lessa_strong--weak_2025,sala_spontaneous_2024,liu_local_2026,divi_local_2026,zhang_local_2026,guo_quantum_2025}, like intrinsically mixed topological phases~\cite{ma_topological_2025,guo_locally_2025,lessa_higher-form_2025,sang_mixed-state_2026,yang_topological_2026}. No observable linear in $\rho$ detects the SW-SSB~\cite{lee_charge_2026,lessa_strong--weak_2025,weinstein_efficient_2025,liu_diagnosing_2025}, but nonlinear diagnostics such as the fidelity correlator do~\cite{wang_strong--weak_2026,weinstein_efficient_2025,lessa_strong--weak_2025,zhang_fluctuation_2024}. SW-SSB is tied to decoding thresholds and the breakdown of quantum memories~\cite{dennis_topological_2002,fan_diagnostics_2024,sang_mixed-state_2024,lee_exact_2025, niwa_coherent_2025, zhang_strong--weak_2025,lessa_strong--weak_2025,sang_stability_2025,vijay_information_2025,kim_error_2026,hauser_information_2026}.

\begin{figure}[t]
\centering
\includegraphics[width=0.95\columnwidth]{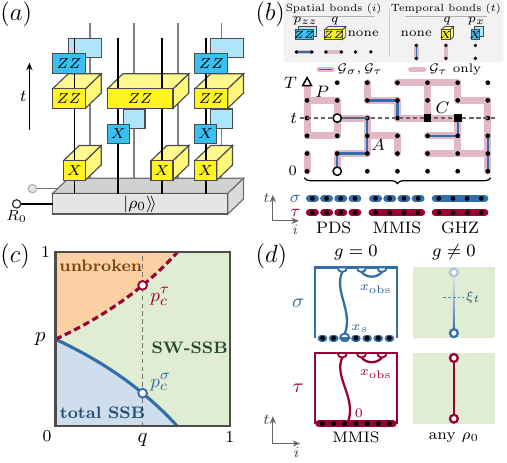}
\caption{\textbf{The monitored circuit maps onto nested percolation graphs with temporal boundary conditions.} \textbf{(a)}~One time step of the doubled circuit on four sites: dephasing and measurement of $X_i$, then of $Z_iZ_{i+1}$. A dephasing (yellow) spans the ket and bra copies, and a measurement (cyan) acts on each. The reference $R_0$ copies $Z_i$ at $t=0$.
\textbf{(b)}~Top: the bond that each event places in $\Gs\subseteq\Gt$. middle: a sample of both graphs, with the endpoints of $A_\alpha$ (circles), $C_\alpha$ (squares) and $P_\alpha$ (triangle). Bottom: free (dashed) and wired (solid) initial rows of the PDS, the MMIS and the GHZ state.
\textbf{(c)}~Phase diagram from \cref{pc} at $k=0.7$, with the critical points at the simulated $q=0.5$ (circles).
\textbf{(d)}~Left: at each critical point ($g=0$), paths of $A_\alpha$ (lines) and $C_\alpha$ (arcs) from the MMIS, with the scaling dimension of each endpoint. Right: in the SW-SSB phase, for any initial state, $\Gs$ is cut beyond the correlation time $\xi_t$ while $\Gt$ still connects.
}
\label{protoc}
\end{figure}

At a conventional critical point, an order parameter relaxes as a power law in time, with an exponent set by the scaling dimension of its field at the initial time~\cite{janssen_new_1989,huse_remanent_1989,zheng_monte_1998,albano_study_2011,yin_universal_2014,lin_probing_2026,wang_relaxation_2026,wang_dynamic_2026,peng_short-time_2026,shu_universal_2026a}. This theory follows a single order parameter, whereas a mixed state has two, one for each symmetry, both visible only to nonlinear diagnostics~\cite{hauser_information_2025}. Studies of SW-SSB have addressed hydrodynamics~\cite{ogunnaike_unifying_2023,hauser_strong--weak_2026,huang_hydrodynamics_2025,gu_spontaneous_2025,sa_exactly_2025}, transitions driven by decoherence~\cite{kim_error_2026,kuno_strong--weak_2024,vijay_information_2025,shu_universal_2026,hauser_strong--weak_2026,haga_discontinuous_2026,chen_strong_2025,fang_algebraic_2026}
or monitoring~\cite{vijay_holographically_2025,agrawal_entanglement_2022,barratt_transitions_2022,singh_mixed-state_2026,hauser_information_2025,zerba_strong--weak_2025}, nonequilibrium steady states~\cite{ziereis_strong--weak_2025,zhang_local_2026}, operator scrambling~\cite{cheng_effective_2026} and experimental
detection~\cite{wang_observation_2026,wei_strong_2026}. How the two order parameters relax at and near an SW-SSB transition has remained open, in particular how the order of the initial state in each of them determines its decay.

In this Letter, we study this critical relaxation in a monitored $\Zt$ circuit [\cref{protoc}(a)], inspired by Ref.~\cite{hauser_information_2025}, with projective measurements and recorded dephasing locations~\cite{li_decodable_2023}. The circuit maps exactly onto two nested bond-percolation problems, one for each order parameter, in which the initial state acts as a temporal boundary condition, wired if the state is ordered in that order parameter and free otherwise. The maximally mixed invariant state (MMIS)~\cite{moharramipour_symmetry_2024}, an SW-SSB state~\cite{weinstein_efficient_2025,ziereis_strong--weak_2025}, wires the strong-symmetry boundary and leaves the weak-symmetry one free, a combination that no pure state realizes.

\begin{figure}[tb]
\centering
\includegraphics[width=\columnwidth]{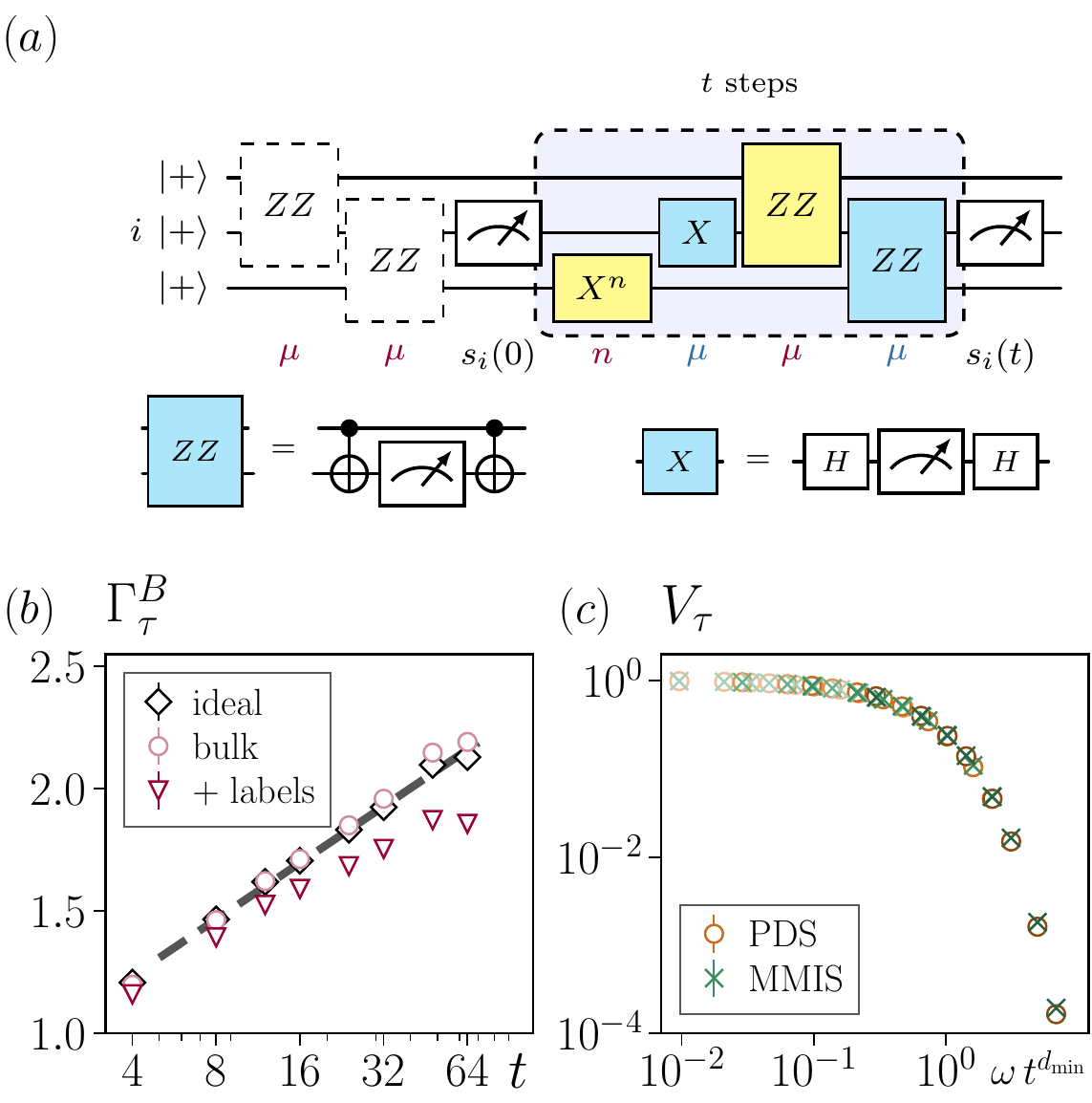}
\caption{\textbf{Environment-assisted decoding measures the temporal boundary condition.}
Labels kept by the environment make every correlator a single-shot average [\cref{expest}].
\textbf{(a)} Circuit on three sites: the round of $ZZ$ measurements that prepares the MMIS (dashed), the readout $s_i(0)$ of site $i$ in place of the reference $R_0$, one of the $t$ steps of \cref{protoc}(a), and the readout $s_i(t)$. Cyan boxes are measurements and yellow boxes unraveled dephasing. Outcomes in blue form the record, and labels in red are stored apart from it and used by the $\tau$ decoder alone. The bottom row shows the circuits that implement the $ZZ$ and $X$ measurements.
\textbf{(b)} Decoded contrast [\cref{expcontrast}] at $p_c^\tau$ and $L=1024$. With bulk errors $\zeta=10^{-2}$ (light red), it stays within $0.06$ of its error-free value (black), which follows the slope $x_s/z$ (dashed). Adding label errors $\zeta_0=10^{-2}$ (dark red) lowers it.
\textbf{(c)} Visibilities [\cref{expvis}] with clean labels for $\zeta=10^{-3}$ to $3\times10^{-2}$ (light to dark). They collapse against $\omega\,t^{d_{\rm min}}$ and nearly coincide for the PDS (orange circles) and the MMIS (green crosses).}
\label{expt}
\end{figure}

Each autocorrelation exponent is the sum of an initial dimension, set by the initial state, and an observation dimension, set by the slice on which it is observed, and the equal-time correlator on the same slice gives the observation dimension alone. For the MMIS, the initial dimension is $1/3$ for the weak-symmetry order parameter and $0$ for the strong-symmetry one, each at its own critical point. The same dimensions set how fast a stored bit is lost, and from the MMIS the system alone loses it as a power law in time while system and environment together retain it. Stabilizer simulations confirm these exponents (\cref{results}). Measuring the nonlinear order parameters directly would require postselection. We provide an experimental protocol to avoid the postselection by keeping an environment label for each dephasing event, which makes every correlator an average over single shots (\cref{expt}), and gate-level noise models indicate that it is feasible on current quantum processors.

\mainsec{sec:model}{Monitored circuit and its two order parameters}We consider $L$ qubits on a ring, with Pauli operators $X_i$ and $Z_i$ at site $i$. In each time step [\cref{protoc}(a)], every $X_i$ is dephased, $\rho\to(\rho+X_i\rho X_i)/2$, with probability $q$ and then measured with probability $p_x=kp$. Every $Z_iZ_{i+1}$ is then dephased with probability $q$ and measured with probability $p_{zz}=k(1-p)$. All events are independent. Here $k$ sets the measurement density and $p$ the fraction of $X_i$ measurements, and we simulate $k=0.7$ and $q=0.5$. Every operation commutes with $\bar X=\prod_iX_i$, a strong $\Zt$ symmetry of the dynamics.

A trajectory is labeled by its record $m$, which contains the locations of all events, dephasing included, and the measurement outcomes~\cite{li_decodable_2023}. In the doubled representation, the normalized state $\rho_m$ becomes the vector $\dket{\rho_m}=\sum_{s,s'}\rho_m(s,s')|s\rangle|s'\rangle$ on a ket and a bra copy~\cite{lee_symmetry_2025,bao_symmetry_2021,hauser_information_2025,ziereis_strong--weak_2025}, where $s_i,s'_i=\pm1$ are the eigenvalues of $Z_i$ on the two copies. The two order parameters are the fields $\sigma_i=s_i$ and $\tau_i=s_is'_i$. The strong symmetry $\bar X\otimes\openone$ flips both, $(\sigma,\tau)\to(-\sigma,-\tau)$, whereas the weak symmetry $\bar X\otimes\bar X$ flips $\sigma$ alone, $(\sigma,\tau)\to(-\sigma,\tau)$~\cite{ziereis_strong--weak_2025}. Long-range order in $\tau$ breaks the strong symmetry and preserves the weak one, while long-range order in $\sigma$ also breaks the weak one. We therefore call $\tau$ the strong-symmetry and $\sigma$ the weak-symmetry order parameter. A state ordered in $\tau$ but disordered in $\sigma$ has SW-SSB.

The $X$-polarized product state (PDS) $|{+}\rangle^{\otimes L}$ is disordered in both fields. The Greenberger--Horne--Zeilinger (GHZ) state $(|{\Uparrow}\rangle+|{\Downarrow}\rangle)/\sqrt2$, with $\Uparrow$ and $\Downarrow$ the fully polarized configurations, is ordered in both. The MMIS $\rho_0=(1+\bar X)/2^L$, the maximally mixed state in the sector $\bar X=1$~\cite{lessa_strong--weak_2025,ziereis_strong--weak_2025}, is ordered in $\tau$ alone and has SW-SSB.

\mainsec{sec:observables}{Correlators, autocorrelations and survival probabilities}We probe $\sigma$ and $\tau$ with correlators quadratic in the trajectory state. For $O=Z_aZ_c$, a product of two charged operators at points $a$ and $c$, the Edwards--Anderson (EA) and R\'enyi-2 correlators are~\cite{hauser_information_2025,lessa_strong--weak_2025}
\begin{equation}
C_\sigma=\overline{(\Tr\rho_mO)^2},\qquad
C_\tau=\overline{\Tr(\rho_mO\rho_mO)/\Tr\rho_m^2},
\label{diag}
\end{equation}
where the overline averages over trajectories with their Born probabilities. $C_{\sigma}$ and $C_{\tau}$ are the correlators of $\sigma$ and of $\tau$, respectively, and the quadratic form in $C_{\sigma}$ removes the record-dependent sign of $\Tr\rho_mO$. In this stabilizer circuit, the R\'enyi-2 correlator equals, on every trajectory, the fidelity correlator that diagnoses SW-SSB~\cite{lessa_strong--weak_2025} (\cref{em:mapping}). The equal-time correlators $C_\alpha(r,t)$, with $\alpha\in\{\sigma,\tau\}$, take $O=Z_iZ_{i+r}$ at time $t$~\cite{ziereis_strong--weak_2025,hauser_information_2025}.

Beyond equal-time correlations, we study single-site autocorrelations $A_\alpha(t;\kappa)$ of $Z_i$ between times $0$ and $t$. To probe these temporal correlations, we introduce reference qubits $R_0$ and $R_1$ that record the $Z$-basis information at site $i$ through controlled-NOT gates at times $0$ and $t$, respectively~\cite{gullans_scalable_2020,liu_entanglement_2024,liu_noisy_2026}, allowing both endpoints to be represented by operators at a common readout time. The system then evolves for an additional interval $\kappa t$, up to $T=(1+\kappa)t$. With $\rho_m$ the joint conditioned state of the system and references at $T$ and $O_R=Z_{R_0}Z_{R_1}$, \cref{diag} gives~\SM{sm:protocol}
\begin{equation}
A_\sigma=\overline{(\Tr\rho_m O_R)^2},\qquad
A_\tau=\overline{\Tr(\rho_m O_R\rho_m O_R)/\Tr\rho_m^2}.
\label{auto}
\end{equation}
Although the endpoints remain at $0$ and $t$, events after $t$ still update the conditioned state. Because \cref{auto} is nonlinear in $\rho_m$, averaging over later outcomes does not in general remove the dependence on $T$. We call the slice at $t$ the insertion slice and set $\kappa=0$ unless stated otherwise. These single-site autocorrelations extend temporal nonlinear correlators defined for global charges~\cite{agrawal_entanglement_2022,barratt_transitions_2022,zerba_strong--weak_2025} or channel dynamics~\cite{gopalakrishnan_lieb_2026}.

The copy at $t=0$ also stores one bit in $R_0$, the initial value of $Z_i$. The survival probability $P_\sigma(t)$ is the probability that an observer holding the record can decode this bit at time $t$ from the system alone, and the bit is then \emph{decodable}. For $P_\tau(t)$, the observer also holds the environment qubits that purify every dephasing event and a mixed initial state, and the bit is \emph{recoverable}~\SM{sm:info}. Thus $A_\alpha$ probes correlations between specified endpoints, whereas $P_\alpha$ tests the recovery of information from the whole accessible output.

\setcounter{topnumber}{1}

\begin{figure}[t]
\centering
\includegraphics[width=\columnwidth]{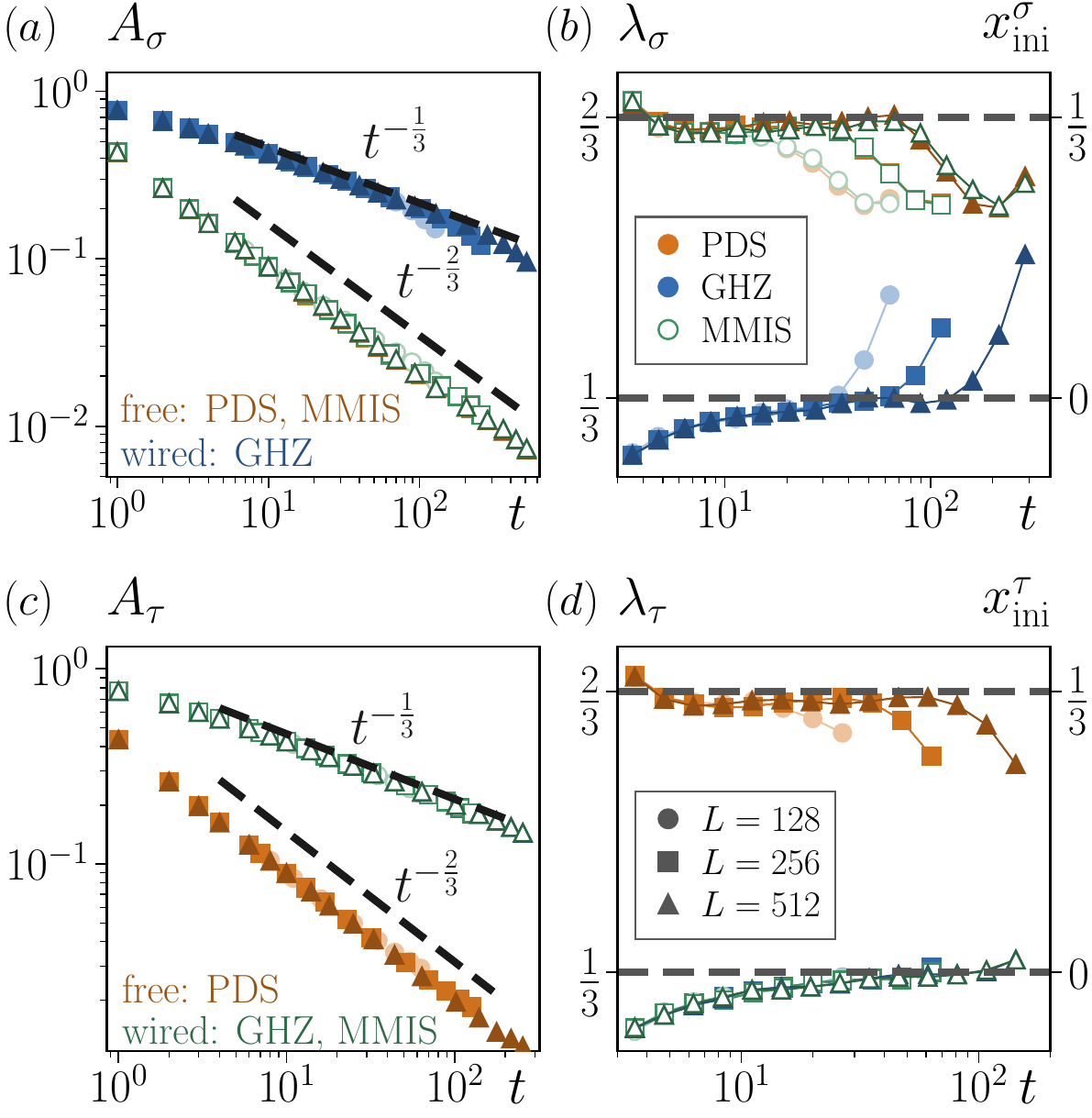}
\caption{\textbf{The MMIS relaxes as the product state in $\sigma$ and as the GHZ state in $\tau$.} Autocorrelations at the critical point of each order parameter from the PDS (orange), the GHZ state (blue) and the MMIS (green, open symbols), for $L=128$, $256$ and $512$.
\textbf{(a)}~$A_\sigma$ at $p_c^\sigma$ and \textbf{(c)}~$A_\tau$ at $p_c^\tau$, with the predicted power laws (dashed).
\textbf{(b,d)}~Local slopes $\lambda_\alpha = -d \ln A_\alpha / d \ln t$ ($z = 1$). The right axes give $\xini^\alpha = \lambda_\alpha - \eta/2$ with the exact $\eta = 2/3$, and the dashed lines mark the predictions $1/3$ (free) and $0$ (wired). The departures at late times, later for larger $L$, are finite-size effects.}
\label{vector}
\end{figure}

\begin{table}[t]
\caption{Autocorrelation exponents $\lambda_\alpha$ and initial dimensions $\xini^\alpha$ for the three initial states, predicted by the exact mapping and simulated at $L=512$ on the final row ($\kappa=0$). Each order parameter is taken at its own critical point. The initial dimensions follow from \cref{master}. \Cref{SMresults} lists further exponents.
}
\label{results}
\begin{ruledtabular}
\begin{tabular}{llllll}
state & & $\lambda_\sigma$ & $\lambda_\tau$ & $\xini^\sigma$ & $\xini^\tau$\\
\hline
PDS & predicted & $2/3$ & $2/3$ & $1/3$ & $1/3$\\
 & simulated & $0.66(1)$ & $0.66(1)$ & $0.33(1)$ & $0.33(1)$\\
MMIS & predicted & $2/3$ & $1/3$ & $1/3$ & $0$\\
 & simulated & $0.66(1)$ & $0.33(1)$ & $0.33(1)$ & $0.00(1)$\\
GHZ & predicted & $1/3$ & $1/3$ & $0$ & $0$\\
 & simulated & $0.33(1)$ & $0.33(1)$ & $0.00(1)$ & $0.00(1)$\\
\end{tabular}
\end{ruledtabular}
\end{table}

\begin{figure}[t]
\centering
\includegraphics[width=\columnwidth]{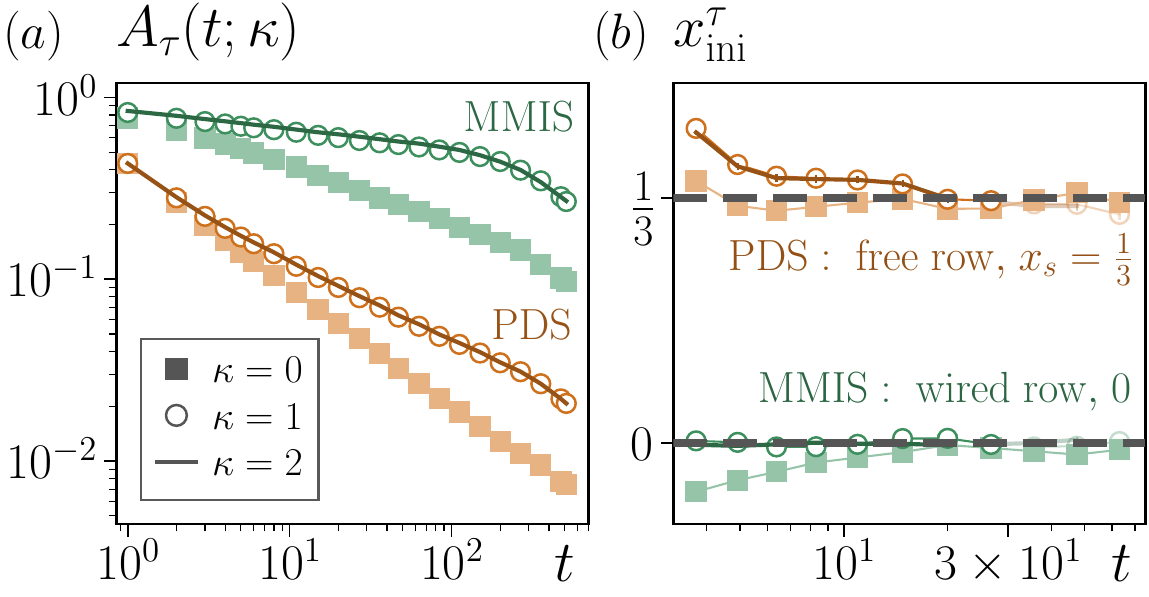}
\caption{\textbf{The initial dimension does not depend on the insertion slice.} Data for $\tau$ at $p_c^\tau$ and $L=512$ from the PDS (orange) and the MMIS (green). The insertion slice is the final row ($\kappa=0$, light filled squares) or lies a time $\kappa t$ below it ($\kappa=1$, open circles; $\kappa=2$, solid lines).
\textbf{(a)}~$A_\tau(t;\kappa)$ decays more slowly for $\kappa>0$.
\textbf{(b)}~$\xini^\tau = \lambda_\tau - \eta_\tau/2$, with $\eta_\tau$ measured on the same slice, approaches $1/3$ for the PDS (free) and $0$ for the MMIS (wired) at every $\kappa$. Data beyond $(1 + \kappa)t = L/8$, where finite size sets in, are drawn lighter.}
\label{slice}
\end{figure}

\begin{figure}[t]
\centering
\includegraphics[width=\columnwidth]{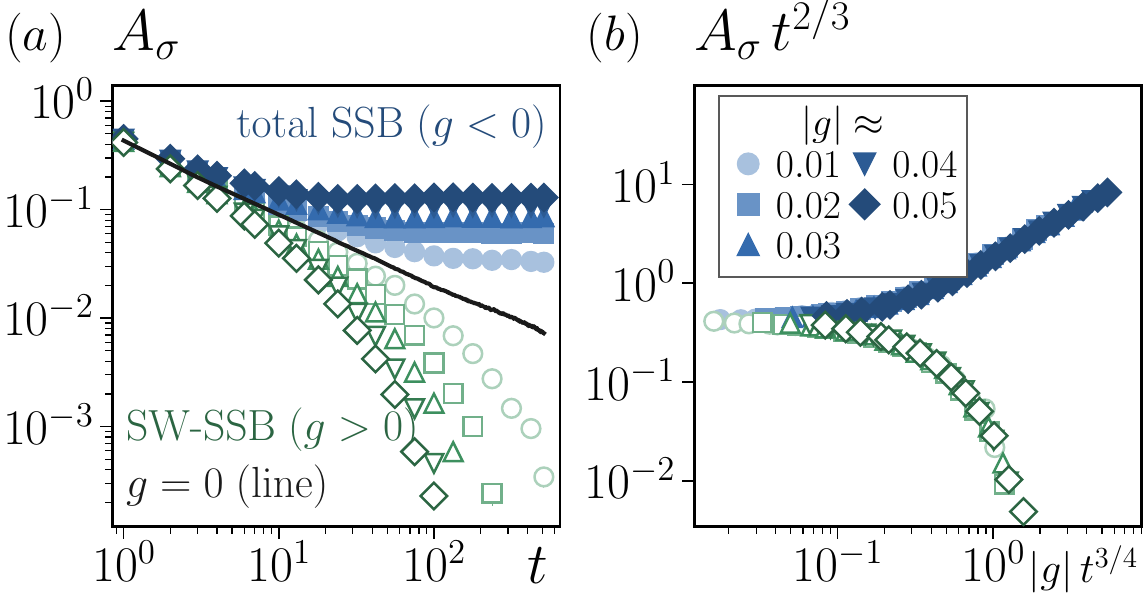}
\caption{\textbf{Relaxation near $p_c^\sigma$ separates total SSB from SW-SSB.} $A_\sigma$ from the PDS at $L = 512$ for $g = p - p_c^\sigma$ with $|g| \approx 0.01, \dots, 0.05$ (pale to deep). Blue filled symbols are $g<0$ (total SSB) and green open symbols $g>0$ (SW-SSB).
\textbf{(a)}~$A_\sigma$ against $t$, with the critical law (black line, $g=0$).
\textbf{(b)}~Collapse of the same data by \cref{scalingA} with the exact $\lambda_\sigma/z = 2/3$ and $1/\nu z = 3/4$, one branch for each sign of $g$.
}
\label{hide}
\end{figure}

\mainsec{sec:mapping}{Exact mapping onto nested percolation}In the $Z$ basis of both copies, the unnormalized trajectory state $\tilde\rho_m\propto\rho_m$ is a sum over configurations of $\sigma$ and $\tau$ on the spacetime lattice of sites $(i,t)$, with one weight per temporal bond from the $X$ events and one per spatial bond from the $ZZ$ events. For the event locations given by the record, each weight factorizes into a function of $\sigma$ and a function of $\tau$, each of which either ties its field across the bond or leaves it free. An untouched temporal bond and a measured $Z_iZ_{i+1}$ tie both fields, a dephasing ties $\tau$ alone, and a measured $X_i$ ties neither (\cref{weight}). The tied bonds thus form two nested graphs $\Gs\subseteq\Gt$ [\cref{protoc}(b)]. Measurement outcomes enter only through signs, which drop out of the quadratic correlators of \cref{diag}. Since the events are independent, each graph is an instance of anisotropic bond percolation, with spatial and temporal bond probabilities $p_h$ and $p_v$ [\cref{bonds}].

For an event pattern $E$, the graph $\mathcal G_\alpha(E;T)$ spans the rows from the initial one at time $0$ to the final one at $T=(1+\kappa)t$. A trajectory contributes $1$ to $C_\alpha$ or $A_\alpha$ if the endpoints are connected ($\leftrightarrow$) in this graph and $0$ otherwise (\cref{em:mapping}). Averaging over $E$ gives
\begin{equation}
\begin{aligned}
C_\alpha(r,t;\kappa)&=\Pr_E[(i,t)\leftrightarrow(i+r,t)],\\
A_\alpha(t;\kappa)&=\Pr_E[(i,0)\leftrightarrow(i,t)].
\end{aligned}
\label{conn}
\end{equation}
The survival probability instead probes the whole final row at $T=t$, $P_\alpha(t)=\Pr_E[(i,0)\leftrightarrow\text{final row}]$.

The order parameter $\alpha$ has long-range order when $C_\alpha$ stays finite at large distance~\cite{lessa_strong--weak_2025}, that is, when $\mathcal G_\alpha$ percolates. Since $\Gs\subseteq\Gt$, $\sigma$ cannot order without $\tau$, and SW-SSB is a percolating $\Gt$ with a non-percolating $\Gs$. The self-duality condition $p_h+p_v=1$ of anisotropic bond percolation gives the critical points exactly~\cite{sykes_exact_1964,kesten_percolation_1982},
\begin{equation}
p_c^\sigma=\frac{k-q}{k(2-q)},\qquad p_c^\tau=\frac{k+q-kq}{k(2-q)},
\label{pc}
\end{equation}
which are $4/21$ and $17/21$ at the simulated parameters. The field $\sigma$ orders for $p<p_c^\sigma$ and $\tau$ for $p<p_c^\tau$, which separates the total-SSB (both ordered), SW-SSB ($\tau$ ordered alone) and unbroken phases [\cref{protoc}(c)].

The critical exponents are those of two-dimensional percolation~\SM{sm:cri}. The correlation-length exponent is $\nu=4/3$~\cite{den_nijs_relation_1979,smirnov_critical_2001}, the order parameter has scaling dimension $x_s=1/3$ at a boundary and $x_b=5/48$ in the bulk~\cite{cardy_conformal_1984,smirnov_critical_2001,nienhuis_magnetic_1980,lawler_one-arm_2002}, and the dynamic exponent is $z=1$, since the anisotropy only rescales time~\cite{li_measurement-driven_2019,lang_entanglement_2020,sang_measurement-protected_2021,sang_entanglement_2021,nahum_entanglement_2020,yu_gapless_2026}.

\mainsec{sec:boundary}{Initial states fix temporal boundary conditions}The initial state fixes the boundary conditions through its matrix elements $\rho_0(s,s')$ on the initial row [\cref{protoc}(b)]. For the three states considered here, order in a field corresponds to a \emph{wired} boundary in its graph, with all initial sites in one cluster, whereas disorder corresponds to a \emph{free} boundary, with no initial bonds. This correspondence follows by expressing $\rho_0(s,s')$ as products of the bond weights of \cref{weight} along that row (\cref{em:initial}). The PDS thus leaves both rows free, the GHZ state wires both, and the MMIS wires the initial row of $\Gt$ alone.

The free-$\sigma$, wired-$\tau$ boundary requires a mixed initial state, because the two correlators in \cref{diag} coincide for pure states. The reverse combination is forbidden (\cref{em:initial}). Free and wired boundaries are boundary fixed points~\cite{janssen_new_1989,zheng_monte_1998,calabrese_critical_2006,yu_nonequilibrium_2026}. The final boundary is free in both graphs (\cref{em:mapping}).

\mainsec{sec:scaling}{Endpoint dimensions determine the decay exponents}At the critical point of $\alpha$ and for $1\ll t\ll L^z$, the probabilities in \cref{conn} are correlation functions in a scale-invariant spacetime slab~\cite{cardy_conformal_1984,janssen_new_1989} [\cref{protoc}(d)]. A connection between endpoints separated by a time $t$ scales as $t^{-(x_1+x_2)/z}$, where $x_1$ and $x_2$ are their scaling dimensions~\SM{sm:auto}. For $A_\alpha$, the initial state fixes the initial dimension $\xini^\alpha$, which is $x_s$ on a free initial row and $0$ on a wired one. The location of the later endpoint fixes the observation dimension $\xobs^\alpha$. It is $x_s$ for $T=t$ ($\kappa=0$), when this endpoint lies on the final boundary, and $x_b$ for fixed $\kappa>0$ in the scaling limit, when it lies in the bulk~\cite{gullans_scalable_2020}. Both endpoints of $C_\alpha$ lie on the insertion slice and have dimension $\xobs^\alpha$, and for $r\ll t^{1/z}$ the initial boundary contributes only corrections~\SM{sm:equal}. With $A_\alpha\propto t^{-\lambda_\alpha/z}$ and $C_\alpha\propto r^{-\eta_\alpha}$, we obtain
\begin{equation}
\lambda_\alpha=\xini^\alpha+\xobs^\alpha,\qquad
\eta_\alpha=2\xobs^\alpha.
\label{laws}
\end{equation}
Combining the two measured exponents then removes $\xobs^\alpha$and isolates the initial dimension (\cref{results}),
\begin{equation}
\xini^\alpha=\lambda_\alpha(\kappa)-\eta_\alpha(\kappa)/2.
\label{master}
\end{equation}
The survival probability $P_\alpha(t)$ instead asks whether the initial endpoint connects anywhere on the final row at $T=t$. Only the initial dimension enters, and $P_\alpha(t)\propto t^{-\xini^\alpha/z}$. The initial dimension thus also sets how fast the stored bit is lost~\SM{sm:info}.  

Away from the critical point, with $g\equiv p-p_c^\alpha$, the correlation time $\xi_t\propto|g|^{-\nu z}$ and the size $L$ enter through a scaling function $\mathcal F_\alpha$~\cite{janssen_new_1989,huse_remanent_1989,zheng_monte_1998},
\begin{equation}
A_\alpha(t,g,L)=t^{-\lambda_\alpha/z}\,\mathcal{F}_\alpha\bigl(g\,t^{1/\nu z},\,t/L^z\bigr).
\label{scalingA}
\end{equation}
The autocorrelation decays exponentially beyond $\xi_t$ in the disordered phase and, with $L\to\infty$ taken first, approaches a plateau $\propto|g|^{\nu\lambda_\alpha}$ in the ordered one~\SM{sm:off}. For a bit stored at one site, $P_\sigma$ stays finite at long times only in the total-SSB phase, and in the SW-SSB phase it decays beyond $\xi_t$ while $P_\tau$ stays finite. Refs.~\cite{li_decodable_2023,hauser_information_2025} found the same phase structure for logical information in the steady state.

\mainsec{sec:numerics}{Simulations confirm the predicted exponents}We simulate the doubled circuit exactly with stabilizer methods for $L\le512$~\SM{sm:protocol}. Error bars in \crefrange{vector}{expt} are one standard error. At the critical point of each order parameter, the autocorrelations decay as $t^{-2/3}$ from a free and as $t^{-1/3}$ from a wired initial row [\cref{vector}(a,c)]. With the exact $\eta=2/3$, the local slopes give $\xini=1/3$ and $0$, respectively [\cref{vector}(b,d), \cref{results}]. The curves for $\sigma$ and $\tau$ differ only in their boundary conditions, because the two graphs have identical bond probabilities at their critical points [\cref{B1}]. The MMIS follows the PDS in $\sigma$ and the GHZ state in $\tau$, and its initial dimensions $(\xini^\sigma,\xini^\tau)=(1/3,0)$, against $(1/3,1/3)$ for the PDS and $(0,0)$ for the GHZ state, are the dynamical fingerprint of an SW-SSB initial state.

Varying the insertion slice changes $\xobs^\tau$ and leaves $\xini^\tau$ unchanged. The measured $\eta_\tau$ is $0.66(1)$ on the final row ($\kappa=0$), against $2x_s=2/3$, and is consistent with $2x_b=5/24$ in the bulk at $\kappa=1,2$ (\cref{Sslice}). Moving the later endpoint of $A_\tau$ into the bulk slows its decay [\cref{slice}(a)], but $\lambda_\tau-\eta_\tau/2$ approaches $1/3$ for the PDS and $0$ for the MMIS at every $\kappa$ [\cref{slice}(b)]~\SM{sm:numerics}.

We sample $P_\alpha$ directly from the percolation graphs for $L\le2048$. At the critical point of $\alpha$, a free initial row, which holds the bit at site $i$ alone, gives $P_\alpha\propto t^{-0.334(2)}$. From a wired row, which shares the bit among all sites, a path from any initial site suffices, and $P_\alpha>0.998$ up to $t=L/8$. The wired constraints of the GHZ state are fixed by the system alone, whereas those of the $\tau$ row of the MMIS require its purifying environment~\SM{sm:info,sm:numerics}. At the two critical points, the MMIS thus keeps the bit recoverable, while the probability $P_\sigma$ of decoding it from the system decays as from the PDS.

Near $p_c^\sigma$, $A_\sigma$ from the PDS saturates to a plateau in the total-SSB phase ($g<0$) and decays exponentially in the SW-SSB phase ($g>0$) [\cref{hide}(a)]. With the exact $\lambda_\sigma/z=2/3$ and $\nu z=4/3$, \cref{scalingA} collapses the data from both sides of the transition without adjustable parameters [\cref{hide}(b)]. A direct fit of the correlation time (\cref{Sxi}) gives $\nu z=1.27(3)$, about two standard errors below $4/3$, a deviation whose origin the present data do not isolate~\SM{sm:numerics}.

\mainsec{sec:experiment}{Measuring the fingerprint without postselection}The correlators of \cref{diag} are quadratic in $\rho_m$, so a direct estimate needs the same record twice, which is exponentially rare. The exact mapping avoids this. Each bond of $\mathcal G_\alpha$ fixes the product of the $Z$ values at its ends to a sign, which the record supplies on the bonds of $\Gs$. The remaining bonds of $\Gt$ come from dephasing events, which we unravel into random operations whose outcomes, the \emph{labels}, are stored apart from the record and used only to decode $\tau$ (\cref{em:proposal}). We read out $Z_i$ at times $0$ and $t$ and, if a path of $\mathcal G_\alpha$ connects the two endpoints, multiply the outcomes by the signs along it. This single-shot estimator $Y_\alpha$, set to zero when there is no path, averages to $A_\alpha$ [\cref{expest}]. Every shot is kept, and the number of shots grows only as $t^{\lambda_\alpha/z}$. Unlike related decoding schemes for other diagnostics of monitored circuits~\cite{li_decodable_2023,agrawal_observing_2023,garratt_probing_2023}, the labels give access to the strong-symmetry order parameter. One round of $Z_iZ_{i+1}$ measurements on $|{+}\rangle^{\otimes L}$, with the outcomes stored as labels, prepares the MMIS exactly [\cref{expt}(a)].

On a processor, errors at unknown locations reduce the decoded average to $B_\alpha=A_\alpha V_\alpha$. For independent Pauli and readout faults, the \emph{visibility} $V_\alpha$ is given exactly by \cref{expvis}. The visibilities of the PDS and the MMIS nearly coincide [\cref{expt}(c)], because the initial state barely changes the length of the connecting paths. They largely cancel in the decoded contrast $\Gamma_\tau^B=\ln(B_\tau^{\rm MMIS}/B_\tau^{\rm PDS})$, which keeps the ideal growth $(x_s/z)\ln t$ set by the difference between the two initial dimensions of $\tau$ [\cref{expcontrast}, \cref{expt}(b)]. Errors on the labels do not cancel and shift the contrast directly. For error rates of current superconducting processors, on which monitored circuits have been realized~\cite{koh_measurement-induced_2023,hoke_measurement-induced_2023,krinner_repeated_2022,google_suppressing_2023}, gate-level noise models keep this contrast resolvable over several time steps at a few times the ideal shot cost (\cref{em:proposal}). Trapped-ion~\cite{noel_measurement-induced_2022,pino_demonstration_2021,moses_race-track_2023} and neutral-atom~\cite{bluvstein_logical_2024} processors offer the same mid-circuit readout.

\mainsec{sec:discussion}{Discussion and outlook}For the initial states studied here, critical relaxation reflects their order through a temporal boundary condition for each order parameter, whose contribution $\lambda_\alpha-\eta_\alpha/2$ isolates [\cref{laws,master}]. For the MMIS, this contribution vanishes for $\tau$ and is positive for $\sigma$, each at its own critical point, a dynamical signature of its SW-SSB order. Each boundary condition is also a storage mode. Retaining the dephasing labels makes these correlators measurable without postselection.

Initial states act as temporal boundary conditions in classical relaxation~\cite{janssen_new_1989,huse_remanent_1989} and in replicated descriptions of open quantum systems~\cite{lee_quantum_2023,tikhanovskaya_universality_2024,li_decodable_2023,zou_channeling_2023,chen_subsystem_2025}, but only a mixed state can impose different ones on the two order parameters. Two-time averages in model A~\cite{hohenberg_theory_1977} do not depend on later histories and obey the standard initial-slip relation~\cite{janssen_phase_1992,calabrese_ageing_2005}~\SM{sm:modelA}. Our nonlinear correlators instead depend on the state conditioned on the record up to $T$, which is therefore a temporal boundary with surface scaling~\cite{gullans_scalable_2020,zabalo_critical_2020,li_conformal_2021,lunt_measurement_2021,hauser_information_2025,sang_entanglement_2021}, as at a spatial boundary~\cite{ritschel_universal_1995,shu_universal_2026a,yu_conformal_2022}.

\Cref{master} may extend beyond percolation when $A_\alpha$ and $C_\alpha$ obey endpoint scaling with the same observation dimension~\SM{sm:scaling}. As a signature of SW-SSB, it also requires the initial-boundary fixed points of the new model, and both conditions are verified only in the present circuit. Weak measurements offer a test, in which the R\'enyi-2 and fidelity correlators need not coincide~\cite{lessa_strong--weak_2025}. Further directions include critical aging~\SM{sm:aging}, other boundary fixed points of mixed initial states~\cite{yang_boundary_2026} and continuous symmetries~\cite{lee_charge_2026,sarma_strong_2026,fang_algebraic_2026}.

\begin{acknowledgments}
\runinhead{Acknowledgments} 
J.-Q. L. and S.Y. are supported by the National Natural Science Foundation of China (Grants No.12675053 and No.12222515), the Research Center for Magnetoelectric Physics of Guangdong Province (Grant No.2024B0303390001), the Guangdong Provincial Key Laboratory of Magnetoelectric Physics and Devices (Grant No.2026B1212170049), and Quantum Science and Technology National Science and Technology Major Project (Grant No.2025ZD0300400). S.L. was supported by the Gordon and Betty Moore Foundation (grant GBMF8685) towards the Princeton theory program, the Gordon and Betty Moore Foundation's EPiQS Initiative (grant GBMF11070), the Global Collaborative Network Grant at Princeton University, the Simons Investigator Grant No. 404513, the NSF MRSEC (grant DMR-2011750), the Simons Collaboration on New Frontiers in Superconductivity (grant SFI-MPS-NFS-00006741-01), the Princeton Catalysis Initiative (PCI), and the Schmidt Transformative Technology Fund at Princeton University.
\end{acknowledgments}

\setcounter{topnumber}{2}
\bibliography{SWSSB}

\clearpage
\onecolumngrid
\pdfbookmark[1]{End Matter}{endmatter}
\begin{center}
{\large\textbf{\textcolor{headcol}{End Matter}}}
\end{center}
\vspace{2pt}
\twocolumngrid

\renewcommand{\theequation}{\theemsec\arabic{equation}}

\emsec{em:mapping}{Bond percolation mapping}Let $\tilde\rho_m$ denote the unnormalized trajectory state. \Cref{diag,auto} use $\rho_m=\tilde\rho_m/\Tr\tilde\rho_m$, and $\Tr\tilde\rho_m$ is the Born probability of the outcomes given the event locations. On the doubled state, a dephasing and a measurement with outcome $\mu=\pm1$ act as
\begin{equation}
\mathcal{D}_O=\tfrac{1}{2}(\openone+O\otimes O),\qquad
\mathcal{M}_{O,\mu}=\Pi_{O,\mu}\otimes\Pi_{O,\mu},
\label{ops}
\end{equation}
with $\Pi_{O,\mu}=\tfrac12(\openone+\mu O)$. Since $\mathcal M_{O,\mu}\mathcal D_O=\mathcal M_{O,\mu}$, a dephasing followed by a measurement of the same operator acts as the measurement alone. Inserting complete sets of $Z$-basis states of both copies after every layer gives
\begin{equation}
\tilde\rho_m(s_T,s'_T)=\sum_{\{s,s'\}_{t<T}}\rho_0(s_0,s'_0)\prod_{\langle a,c\rangle}W_{ac},
\label{ZP}
\end{equation}
where $s_t$ is the configuration of row $t$ and $\langle a,c\rangle$ runs over the bonds. The $X$ layers of step $t$ give one weight $W_{ac}$ per temporal bond between rows $t-1$ and $t$, and the $ZZ$ layers give one weight per spatial bond of row $t$. Each weight depends on the fields only through $u=\sigma_a\sigma_c=s_as_c$ and $v=\tau_a\tau_c$, with $s'_as'_c=uv$, and each of the six weights in \cref{weight} is a product $W_{ac}=f(u)\,h(v)$. A factor $f\propto1+\mu u$, with $\mu=1$ on the untouched temporal bond, enforces $\sigma_a\sigma_c=\mu$ and puts the bond in $\Gs$. A factor $h\propto1+v$ enforces $\tau_a=\tau_c$ and puts it in $\Gt$. A constant or the sign $v^{(1-\mu)/2}$ imposes no constraint, although the sign still enters the weight. With the event probabilities of the circuit, \cref{weight} gives the spatial and temporal bond probabilities
\begin{equation}
\begin{aligned}
\Gs:&\quad p_h=p_{zz}, &&p_v=(1-p_x)(1-q),\\
\Gt:&\quad p_h=1-(1-p_{zz})(1-q), &&p_v=1-p_x,
\end{aligned}
\label{bonds}
\end{equation}
and $\Gs\subseteq\Gt$ because $1+\mu u$ never occurs without $1+v$. Without $ZZ$ dephasing, $\Gs$ is the set of connected bonds of Ref.~\cite{li_decodable_2023} and $\Gt$ the complement of its broken bonds. A reference qubit carries the fields of the vertex it copies~\SM{sm:protocol}.

\begin{table}[b]
\caption{Bond weights for a measurement (meas.) with outcome $\mu$, a dephasing (deph.), or no operation, along with the resulting bonds in the two graphs.}
\label{weight}
\begin{ruledtabular}
\begin{tabular}{lllcc}
bond & event & $W_{ac}$ & $\Gs$ & $\Gt$\\
\hline
spatial & $ZZ$ meas., $\mu$ & $\tfrac{1}{4}(1+\mu u)(1+v)$ & \ding{51} & \ding{51}\\
 & $ZZ$ deph. & $\tfrac{1}{2}(1+v)$ & \ding{55} & \ding{51}\\
 & none & $1$ & \ding{55} & \ding{55}\\
temporal & none & $\tfrac{1}{4}(1+u)(1+v)$ & \ding{51} & \ding{51}\\
 & $X$ deph. & $\tfrac{1}{4}(1+v)$ & \ding{55} & \ding{51}\\
 & $X$ meas., $\mu$ & $\tfrac{1}{4}\,v^{(1-\mu)/2}$ & \ding{55} & \ding{55}\\
\end{tabular}
\end{ruledtabular}
\end{table}

The measurement outcomes enter only through signs. A record whose signs are inconsistent around a loop of $\Gs$, or which leaves an odd number of sign insertions on a cluster of $\Gt$ that misses the final row, has zero Born weight. The remaining signs are pure gauge and cancel in the quadratic correlators~\SM{sm:sign}.

The final row is free in both graphs. Because $\tilde\rho_m$ is proportional to a stabilizer projector, its nonzero matrix elements in the $Z$ basis have equal magnitude. The EA and R\'enyi-2 normalizations, $(\Tr\tilde\rho_m)^2$ and $\Tr\tilde\rho_m^2$, are then equal-weight sums over the allowed configurations, which leaves $\sigma$ free on the final row in the EA correlator and $\tau$ free in the R\'enyi-2 correlator. For every record that occurs, the summand of each correlator is $1$ if its endpoints are connected in $\mathcal G_\alpha$ and $0$ otherwise. It depends on the event pattern alone, so the Born average reduces to the average over event patterns in \cref{conn}. For a Pauli operator $O$, $O\tilde\rho_mO$ either equals $\tilde\rho_m$ or is orthogonal to it, so the R\'enyi-2 and fidelity correlators coincide on every trajectory~\cite{lessa_strong--weak_2025,divi_local_2026}.

\emsec{em:initial}{Initial states as temporal boundary conditions}The initial state enters \cref{ZP} through its matrix elements $\rho_0(s,s')$. The PDS $|{+}\rangle^{\otimes L}$ has $\rho_0(s,s')=2^{-L}$ for all $s$ and $s'$. Since $\bar X|s\rangle=|-s\rangle$, the MMIS $(1+\bar X)/2^{L}$ has $\rho_0(s,s')=2^{-L}(\delta_{s,s'}+\delta_{s',-s})$, which is nonzero only if $\tau_j=s_js'_j$ takes the same value $\pm1$ on every site. The matrix elements of the GHZ state are nonzero only if, in addition, $\sigma_j=s_j$ is uniform. With $u_j=\sigma_j\sigma_{j+1}$ and $v_j=\tau_j\tau_{j+1}$ on the initial row, the three states read
\begin{equation}
\rho_0(s,s')\propto\prod_j
\begin{cases}
1 & \text{PDS},\\
(1+v_j) & \text{MMIS},\\
(1+u_j)(1+v_j) & \text{GHZ}.
\end{cases}
\end{equation}
The factors $1+u_j$ and $1+v_j$ wire the $\sigma$ and $\tau$ rows, respectively. No state wires $\sigma$ and leaves $\tau$ free. If $\rho_0(s,s')$ vanishes unless $s=\pm s^*$ for a fixed configuration $s^*$, so that $\sigma$ is wired, hermiticity $\rho_0(s,s')=\rho_0(s',s)^*$ also requires $s'=\pm s^*$. Then $\tau_j=s_js'_j$ is uniform, and $\tau$ is wired too. The three cases thus exhaust the possible pairs of boundary conditions, and a free $\sigma$ with a wired $\tau$ requires a mixed state (\cref{sec:boundary}). A weight that does not factorize, as for a tilted product state, adds boundary couplings. We expect them to flow to the free or the wired fixed point according to whether the state is disordered or ordered in that field, as an initial magnetization does in model A~\cite{janssen_new_1989}.

\emsec{em:proposal}{Environment-assisted decoding protocol}To probe the temporal boundary conditions, we compare $A_\tau(t)$ for the PDS and the MMIS [\cref{expt}(a)]. The PDS is $|{+}\rangle^{\otimes L}$. One round of $ZZ$ measurements on it leaves a GHZ state in a frame fixed by the outcomes. Averaged over these outcomes, which the environment keeps, this state is exactly the MMIS~\SM{sm:exp-prep}. After the preparation, we measure $Z_i$ in place of the reference qubit $R_0$, which gives $s_i(0)$, run the circuit for $t$ steps, and measure $Z_i$ again to obtain $s_i(t)$. During the evolution we keep the usual record and \emph{unravel} each dephasing into a \emph{label} kept by the environment, that is, stored apart from the record. A dephased $Z_iZ_{i+1}$ is measured with its outcome withheld from the record, and a dephased $X_i$ is replaced by $X_i^{n}$ with a fair random bit $n$. Discarding the labels restores the dephasing channels.

Each bond imposes $s_as_c=\mu_{ac}$, with the sign supplied by the record on $\Gs$ and by the labels on the additional bonds of $\Gt$. If $(i,0)$ and $(i,t)$ are connected in $\mathcal G_\alpha$, we choose a path $\mathcal P$ between them without using the outcomes and set $Y_\alpha=\mu_{\mathcal P}s_i(0)s_i(t)$, with $\mu_{\mathcal P}$ the product of the signs along $\mathcal P$. Otherwise $Y_\alpha=0$. In the ideal circuit, the bond constraints give $s_i(0)s_i(t)=\mu_{\mathcal P}$, so that $Y_\alpha=1$ on every connected pattern, and by \cref{conn}
\begin{equation}
A_\alpha(t)=\overline{Y_\alpha}.
\label{expest}
\end{equation}
A relative rms error $\delta$ requires $N\simeq1/(\delta^2A_\alpha)\propto t^{\lambda_\alpha/z}$ shots~\SM{sm:exp-est}.

On a processor, errors at unknown locations flip some of these signs. For a fixed decoder, let Pauli faults and readout flips $f$ occur independently given the event pattern, with state-independent probabilities $0\le\zeta_f<\frac12$, and let $A_\alpha>0$. The \emph{decoded correlator} $B_\alpha$, the average of $Y_\alpha$ with errors, then obeys
\begin{equation}
B_\alpha=A_\alpha V_\alpha,\qquad V_\alpha=\bigl\langle e^{-\Omega_\alpha}\bigr\rangle_{\rm conn},
\label{expvis}
\end{equation}
where $V_\alpha$ is the \emph{visibility} and $\langle\cdot\rangle_{\rm conn}$ averages over connected patterns. The exponent $\Omega_\alpha=\sum_f\chi_f\omega_f$, with $\omega_f=-\ln(1-2\zeta_f)$, sums over the faults that flip the sign of $Y_\alpha$, for which $\chi_f=1$~\SM{sm:exp-err}.

For independent bulk bond errors of probability $\zeta$, clean labels and ideal endpoints, $\Omega_\alpha=\omega|\mathcal P|$, with $\omega=-\ln(1-2\zeta)$ and $|\mathcal P|$ the number of bulk bonds on the path. For the shortest-path decoder, the mean path length $\langle|\mathcal P|\rangle$ grows as $t^{1.13}$ at criticality, consistent with the shortest-path dimension $d_{\rm min}=1.1308$ of percolation~\cite{grassberger_spreading_1992,zhou_shortest-path_2012}, and the sampled visibilities collapse as $V_\alpha\simeq\mathcal V_\alpha(\omega\,t^{d_{\rm min}})$ [\cref{expt}(c)]. They nearly coincide for the PDS and the MMIS, because the initial state decides which patterns connect but hardly how long the connecting paths are. To leading order at $p_c^\tau$ and $\kappa=0$, \cref{laws} then gives the decoded contrast
\begin{equation}
\Gamma_\tau^B\equiv\ln\frac{B_\tau^{\rm MMIS}}{B_\tau^{\rm PDS}}=\frac{x_s}{z}\ln t+\text{const}+\ln\frac{V_\tau^{\rm MMIS}}{V_\tau^{\rm PDS}}.
\label{expcontrast}
\end{equation}
The observation dimension $\xobs^\tau$ cancels in this ratio, and the ideal slope $x_s/z$ is the difference between the two initial dimensions of $\tau$. With clean labels, bulk errors of probability $10^{-2}$ shift $\Gamma_\tau^B$ by at most about $0.06$ up to $t=64$, whereas the error-free contrast rises by $0.66$ between $t=8$ and $64$. Bulk errors thus mainly raise the shot cost, whereas label errors do not cancel and shift the contrast directly [\cref{expt}(b)].

At the gate level, we compile the measurement circuits of \cref{expt}(a) into controlled-$Z$ gates and mid-circuit readout. Error parameters typical of current superconducting processors~\cite{krinner_repeated_2022,google_suppressing_2023} give bond error probabilities of about $(1\text{--}3)\times10^{-2}$ when the faults are pooled into independent flips of each bond. Simulations of up to nine qubits with amplitude damping and asymmetric readout deviate from \cref{expvis} for the matched Pauli noise by at most about $3\times10^{-3}$~\SM{sm:exp-gate}. At $p_c^\tau$, this pooled model gives $V_\tau\simeq\frac12$ at $t=8$ on large rings, at roughly $3$ to $7$ times the ideal shot cost. Within this model, the MMIS boundary condition is testable on current processors over several time steps.

\clearpage
\onecolumngrid
 
\begin{center}
\textbf{\large Supplemental Material}\\[2mm]
\textit{Relaxation critical dynamics of strong-to-weak spontaneous symmetry breaking}\\[1mm]
Jiaqiang Li, Xianquan Yan, Han-Ze Li, Shuai Yin, and Shuo Liu
\end{center}
 
\vspace{4mm}
 
\setcounter{equation}{0}
\setcounter{figure}{0}
\setcounter{table}{0}
\setcounter{section}{0}
\renewcommand{\theequation}{S\arabic{equation}}
\renewcommand{\thefigure}{S\arabic{figure}}
\renewcommand{\thetable}{S\arabic{table}}
\renewcommand{\thesection}{S\arabic{section}}
\setcounter{secnumdepth}{2}
\renewcommand{\thesubsection}{\Alph{subsection}}
The End Matter gives the bond weights, the initial boundary conditions and the decoding protocol (\crefrange{em:mapping}{em:proposal}). This Supplemental Material contains the following sections.
 
\begin{center}
\begin{minipage}{0.8\textwidth}
\small%
\newcommand{\tocS}[2]{\noindent\makebox[2.8em][l]{\ref{#2}}{\hypersetup{linkcolor=black}\hyperref[#2]{#1}}\dotfill\makebox[1.4em][r]{\pageref{#2}}\par}%
\newcommand{\tocA}[3]{\noindent\hspace*{2.8em}\makebox[1.6em][l]{\hyperref[#3]{#1}}{\hypersetup{linkcolor=black}\hyperref[#3]{#2}}\dotfill\makebox[1.4em][r]{\pageref{#3}}\par}%
\tocS{Signs and the final-time boundary}{sm:sign}
\tocS{Critical points and exponents}{sm:cri}
\tocS{Information recoverability}{sm:info}
\tocS{Verification of the mapping}{sm:verify}
\tocS{Protocol and algorithm}{sm:protocol}
\tocS{Scaling details}{sm:scaling}
\tocA{A}{Autocorrelation}{sm:auto}
\tocA{B}{Equal-time correlator}{sm:equal}
\tocA{C}{Insertion inside the bulk}{sm:bulk}
\tocA{D}{Relaxational dynamics}{sm:modelA}
\tocA{E}{Waiting time}{sm:aging}
\tocS{Away from the transition}{sm:off}
\tocS{Numerical procedures and errors}{sm:numerics}
\tocS{Finite-size regime}{sm:z}
\tocS{Environment-assisted decoding protocol}{sm:exp}
\tocA{A}{Single-shot estimators without postselection}{sm:exp-est}
\tocA{B}{Mixed initial state from one round of measurements}{sm:exp-prep}
\tocA{C}{Exact attenuation by unrecorded errors}{sm:exp-err}
\tocA{D}{Gate-level implementation and feasibility}{sm:exp-gate}
\end{minipage}
\end{center}

\section{Signs and the final-time boundary}
\label{sm:sign}
This section shows that the measurement outcomes enter the mapping of \cref{em:mapping} only through pure-gauge signs and that the final row is free in both graphs. As there, $\tilde\rho_m$ is an unnormalized trajectory state and \cref{ZP} its expansion in bond weights. Under the Choi--Jamio\l{}kowski isomorphism, a dephasing is a projector on the doubled Hilbert space, and a measurement acts identically on both copies with the same outcome, which preserves the trajectory probabilities exactly~\citep{bao_symmetry_2021,ziereis_strong--weak_2025}.

The outcomes enter only through signs, and the Born rule restricts these signs to a pure gauge. For $\sigma$, the constraints $\sigma_a\sigma_c=\mu_{ac}$ must hold around every loop $\gamma$ of $\Gs$, so a record with $\prod_{\gamma}\mu_{ac}=-1$ has zero Born weight. Every record that occurs is pure gauge, $\mu_{ac}=\epsilon_a\epsilon_c$ with $\epsilon_a=\pm1$, and
\begin{equation}
\frac{\Tr(\tilde\rho_mZ_aZ_c)}{\Tr\tilde\rho_m}=
\begin{cases}\epsilon_a\epsilon_c, & a\leftrightarrow c\ \text{in}\ \Gs,\\ 0, & \text{otherwise}.\end{cases}
\end{equation} 
If $a$ and $c$ are not connected, reversing $\sigma$ on the cluster of $a$ leaves every weight unchanged and reverses $Z_aZ_c$, so that its expectation value vanishes.
For $\tau$, the field is uniform on each
cluster of $\Gt$. An $X$ measurement with $\mu=-1$ inserts the sign $v$ across a bond outside $\Gt$. Summing over $\tau$ on any cluster that does not reach the final row then forbids an odd number of insertion ends on that cluster. On a cluster that does reach the final row, the insertions give each matrix element a sign that drops out of $|\tilde\rho_m(s_T,s'_T)|^2$ and is $+1$ on the diagonal. This cancellation of signs in quadratic correlators is known from Refs.~\cite{li_decodable_2023,kuno_strong--weak_2024,kuno_intrinsic_2025}.

The final-time boundary follows from the normalizations in \cref{diag},
\begin{equation}
\begin{aligned}
\frac{\Tr\tilde\rho_mO}{\Tr\tilde\rho_m}&=\frac{\sum_s\tilde\rho_m(s,s)O(s)}{\sum_s\tilde\rho_m(s,s)},\\
\frac{\Tr\tilde\rho_mO\tilde\rho_mO}{\Tr\tilde\rho_m^2}&=\frac{\sum_{s,s'}|\tilde\rho_m(s,s')|^2O(s)O(s')}{\sum_{s,s'}|\tilde\rho_m(s,s')|^2},
\end{aligned}
\end{equation}
with $s=s_T$, $s'=s'_T$ and $O(s)$ the eigenvalue of a $Z$-type $O$. Because $|\tilde\rho_m(s,s')|$ takes a single nonzero value for the stabilizer states here, both denominators sum with equal weight over final-row configurations compatible with the constraints. The final row is therefore free for $\sigma$ in the EA correlator and for $\tau$ in the R\'enyi-2 correlator, which gives the connection probabilities of \cref{conn} (\cref{em:mapping}).

\section{Critical points and exponents}
\label{sm:cri}
Bond percolation on the square lattice with probabilities $(p_h,p_v)$ is dual to percolation with $(1-p_v,1-p_h)$ and is critical on the self-dual line $p_h+p_v=1$~\cite{sykes_exact_1964,kesten_percolation_1982}.
Inserting \cref{bonds} gives \cref{pc}. The mapping is exact for every $q$. For $0<q<k$ both critical points lie inside $0<p<1$. At $q=0$ they merge at $1/2$, where the SW-SSB phase closes, and at $q=k$ they reach $p_c^{\sigma}=0$ and $p_c^{\tau}=1$, so that for $q\geq k$ every $0<p<1$ lies in the SW-SSB phase [\cref{protoc}(c)].

Without $ZZ$ dephasing, $\Gs$ and the complement of $\Gt$ are the connected and broken bonds of Ref.~\cite{li_decodable_2023}. The $ZZ$ dephasing adds spatial bonds to $\Gt$ alone and makes the two graphs mutually dual under $p\to1-p$. This map exchanges $p_x$ and $p_{zz}$, and \cref{bonds} then maps $\Gt$ at $1-p$ onto the planar dual of $\Gs$ at $p$. This is the Kramers--Wannier duality of Ref.~\cite{hauser_information_2025}, and it gives $p_c^\sigma+p_c^\tau=1$. The total-SSB, SW-SSB and unbroken phases are the spin-glass, trivial and paramagnetic phases of Ref.~\cite{li_decodable_2023}, as identified in Ref.~\cite{hauser_information_2025}.

At its critical point each graph has
\begin{equation}
(p_h,p_v)=\Bigl(\frac{k+q-kq}{2-q},\,\frac{(2-k)(1-q)}{2-q}\Bigr),
\label{B1}
\end{equation}
$(17/30,13/30)$ at our working point, so that $\sigma$ at $p_c^\sigma$ and $\tau$ at $p_c^\tau$ have identical critical correlators for equal boundary conditions. The exponents are those of two-dimensional percolation, with $\nu=4/3$ and the order-parameter exponents $\beta=5/36$ in the bulk and $\beta_1=4/9$ at a free boundary, which give $x_b=\beta/\nu=5/48$ and $x_s=\beta_1/\nu=1/3$~\cite{den_nijs_relation_1979,nienhuis_magnetic_1980,cardy_conformal_1984,smirnov_critical_2001,lawler_one-arm_2002}. These values are proven for site percolation on the triangular lattice and hold for the present lattice by universality. The anisotropy rescales time relative to space by a constant factor, so $z=1$.

\emph{Other working points.} We repeated the relaxation at $(k,q)=(0.9,0.6)$ and $(0.5,0.2)$, where \cref{B1} gives $(p_h,p_v)=(24/35,11/35)$ and $(1/3,2/3)$. Since only connection probabilities enter, we sampled the bond-percolation graphs of \cref{bonds} directly (\cref{sm:numerics}), on rings of $L=4096$ with $32768$ trajectories averaged over the sites and errors from the scatter of $64$ independent batches. Fits of $\ln A$ against $\ln t$ over $36\le t\le512$ give $\lambda=0.663(1)$ and $0.662(1)$ from a free initial row and $0.331(1)$ and $0.329(1)$ from a wired one, against $0.663(1)$ and $0.330(1)$ at $(0.7,0.5)$. Over $200\le t\le512$ all six values agree with $2/3$ and $1/3$ within two standard errors. On the final row at $t=512$, the slope of $\ln C$ against $\ln r$ over $11\le r\le22$ gives $\eta=0.665(1)$ and $0.669(1)$, against $0.666(1)$ at $(0.7,0.5)$ and the exact $2/3$. Over the shorter distances $5\le r\le11$ the lattice correction depends on the anisotropy, and the same fit gives $0.64$ and $0.66$, against $0.66$.

\section{Information recoverability}
\label{sm:info}
This section relates the survival probabilities to the two graphs and, for the MMIS, to the reference entropy of \cref{sm:z}. Each dephasing channel can be viewed as a unitary coupling of the system to a purifying environment that is subsequently traced out~\cite{weinstein_efficient_2025}. With this purification, the stabilizer formalism relates the bit stored in $R_0$ to the two graphs. A dephasing of $X_j$ prepares an environment qubit $E$ in $|+\rangle$ and applies a controlled-NOT from $E$ to $j$. A dephasing of $Z_jZ_{j+1}$ prepares $E$ in $|0\rangle$ and applies controlled-NOTs from $j$ and $j+1$ to $E$, which thereby records $Z_jZ_{j+1}$. Further environment qubits purify a mixed $\rho_0$. System, references and environment form a pure stabilizer state with a generating set consisting of products of $X$ operators alone or of $Z$ operators alone, the Calderbank--Shor--Steane (CSS) form. Controlled-NOT gates and measurements of these Pauli products preserve this form.

The bit is tracked by the subgroup $\mathcal S_Z$ of $Z$-type stabilizers, generated at $t=0$ by $Z_{R_0}Z_i$ and the $Z$-type generators of the purified $\rho_0$. With $\mathcal{D}_O$ and $\mathcal{M}_O$ the dephasing and the measurement of $O$, the group evolves as 
\begin{equation}
\begin{aligned}
\mathcal D_{X_j}&:\ Z_j\to Z_jZ_E, & \mathcal D_{Z_jZ_{j+1}}&:\ +Z_jZ_{j+1}Z_E,\\
\mathcal M_{Z_jZ_{j+1}}&:\ +Z_jZ_{j+1}, & \mathcal M_{X_j}&:\ -Z_j.
\end{aligned}
\end{equation}
Here the arrow replaces $Z_j$ in every element, a plus sign adds a generator, and a minus sign removes every element containing $Z_j$. Elements containing $Z_{R_0}$ therefore trace paths of bonds from $(i,0)$. Untouched temporal bonds and $ZZ$ measurements, the bonds of $\Gs$, carry $Z_j$ up or sideways without an environment factor. The other bonds of $\Gt$ and a passage through the initial row of a mixed state attach a factor $Z_E$. In a stabilizer state, $Z_{R_0}$ is either fully correlated with a subsystem or uncorrelated with it, so that, with $Q$ a Pauli string,
\begin{equation}
\begin{aligned}
P_\sigma&=\Pr\bigl[\exists\,Z_{R_0}Q\in\mathcal S_Z,\ Q\ \text{on the system}\bigr],\\
P_\tau&=\Pr\bigl[\exists\,Z_{R_0}Q\in\mathcal S_Z,\ Q\ \text{on system and environment}\bigr],
\end{aligned}
\end{equation}
which are the probabilities that $(i,0)$ reaches the final row in $\Gs$ and in $\Gt$, respectively, as given after \cref{conn}. The nesting gives $P_\sigma\le P_\tau$. With probability $P_\tau-P_\sigma$ the bit is \emph{hidden}, that is, recoverable but not decodable, and with probability $1-P_\tau$ the $X$ measurements have erased it. The record carries no information on the bit, because every element containing $Z_{R_0}$ has an odd number of system $Z$ operators while every measured operator has an even number.

An ordered initial state fixes every product $Z_iZ_j$ on the initial row, in the system alone for the GHZ state and in system and purifying environment together for the MMIS. The bit is then shared by all sites, and from a wired row $(i,0)$ reaches the final row whenever any site of that row does. By planar duality on the cylinder,
\begin{align}
1-P_{\alpha}^{\text{wired}}=\Pr[\text{winding dual path through bonds}\notin \mathcal{G}_\alpha],
\end{align}
which for $\Gt$ means that the recorded $X$ measurements, linked by stretches without $ZZ$ events, jointly measure $\bar X$. This measurement determines the state of the reference qubit that is entangled with $\bar X$ in Ref.~\cite{hauser_information_2025} (\cref{sm:z}). From the MMIS, $P_\tau$ therefore equals the entropy $S_R$ of that reference qubit. This probability equals one for $L\to\infty$ and decays on the scale $L^z$ at $p_c^\tau$. In the unbroken phase, with $\xi\propto|g|^{-\nu}$ the correlation length, it becomes small only for $t\gg \xi_t\ln(L/\xi)$, whereas in the SW-SSB phase it stays close to one for times exponential in $L/\xi$. We verified \cref{conn} for $\kappa\ge0$ and the identity $P_\tau=S_R$ trajectory by trajectory on small systems (\cref{sm:verify}), and \cref{sm:numerics} gives $P_\alpha$ for $L$ up to $2048$.

\section{Verification of the mapping}
\label{sm:verify}
\emph{Exact check on small systems.} \Cref{em:mapping} and \crefrange{sm:sign}{sm:info} derive the bond-percolation mapping and the information recoverability in the doubled representation, using sign arguments. We check these results by an independent exact computation in the original Hilbert space, which uses neither the doubled representation nor the sign arguments.

The computation keeps the global state pure. As in \cref{sm:info}, every dephasing event is replaced by a controlled-NOT coupling to a fresh environment qubit $E$, which is never measured. The MMIS is already mixed at $t=0$. For it we add $L$ environment qubits at the start, prepared together with the system in a pure state whose reduced state on the system is the MMIS. System, reference qubits and environment then stay in a pure stabilizer state at all times. A stabilizer tableau, which stores a set of Pauli generators of this state, evolves it exactly. We do not track the signs of the generators. Every statement checked in \cref{smcheck} asks only whether an operator belongs to the stabilizer group up to a sign, and the answer is therefore the same for every set of measurement outcomes.

The events of each trajectory are drawn with random $p$, $q$ and $k$. The two graphs $\Gs$ and $\Gt$ are built from the same list of events, so that each check compares the quantum computation with the graphs of the same trajectory. Three reference qubits copy local $Z$ operators by controlled-NOT gates. The qubit $R_0$ copies $Z_i$ at $t=0$, and $R_1$ and $R_2$ copy $Z_i$ and $Z_j$, respectively, at a time $t_1\le T$ that defines the insertion slice. The pair $R_0$, $R_1$ gives the autocorrelation, and the pair $R_1$, $R_2$ gives an equal-time correlator on that slice. We write $[\cdot]=1$ if a statement holds and 0 otherwise, $\mathcal{S}$ for the stabilizer group of the global state and $\mathcal{S}'$ for that of the reduced state of system and references. In every trajectory we then compare the quantum statements of \cref{smcheck} with connections in the graphs,
\begin{equation}
\begin{aligned}
&\bigl[\exists\,Z_{R_0}Q\in\mathcal S,\ Q\ \text{on the system}\bigr]
=\bigl[(i,0)\leftrightarrow\text{final row in }\Gs\bigr],\\
&\bigl[\exists\,Z_{R_0}Q\in\mathcal S,\ Q\ \text{on system and environment}\bigr]
=\bigl[(i,0)\leftrightarrow\text{final row in }\Gt\bigr],\\
&\bigl[\pm O_{ac}\in\mathcal S'\bigr]=\bigl[a\leftrightarrow c\ \text{in }\Gs\bigr],\qquad
\bigl[O_{ac}\ \text{commutes with}\ \mathcal S'\bigr]=\bigl[a\leftrightarrow c\ \text{in }\Gt\bigr],
\end{aligned}
\label{smcheck}
\end{equation}
for $O_{ac}=Z_{R_0}Z_l$ with $(a,c)=((i,0),(l,T))$ and every final-row site $l$, for $O_{ac}=Z_{R_0}Z_{R_1}$ with $(a,c)=((i,0),(i,t_1))$, and for $O_{ac}=Z_{R_1}Z_{R_2}$ with $(a,c)=((i,t_1),(j,t_1))$, all in the slab of height $T$. The first two lines of \cref{smcheck} test the survival probabilities. They state that the bit in $R_0$ can be read from the system alone, or from system and environment together, exactly when $(i,0)$ is connected to the final row in $\Gs$ or in $\Gt$, respectively. The last line tests the correlators, because in a stabilizer state
\begin{equation}
(\Tr\rho O)^2=\bigl[\pm O\in\mathcal S'\bigr],\qquad
\frac{\Tr\rho O\rho O}{\Tr\rho^2}=\bigl[O\ \text{commutes with}\ \mathcal S'\bigr],
\end{equation}
so that it gives the correlators of \cref{diag} for $C_\alpha$ and $A_\alpha$ in one trajectory, with $\kappa=T/t_1-1\ge0$. Since no sign enters, each comparison holds for every record that occurs, whatever its probability.

For $L=2,\dots,6$, $T=1,2,3,5,6$, the PDS, the MMIS and the GHZ state, and $12000$ trajectories, we find no mismatch. All three fates of the bit (decodable, hidden, erased) occur in at least $18\%$ of the trajectories of each initial state.

We also check that the record alone cannot reveal the value stored in $R_0$. In $15000$ further trajectories, we check before each of the $13157$ measurements of an operator $O$ that $Z_{R_0}O\notin \mathcal{S}$, so that no outcome is correlated with the bit.

Finally, we test the identification of $P_\tau$ with the reference entropy $S_R$ of Ref.~\cite{hauser_information_2025}. In the state of \cref{sm:z}, a reference qubit is entangled with the global charge $\bar X$, and $S_R$ is its entropy after the evolution. In $1500$ trajectories of this state, with the same $L$ and $T$ and random $p$, $q$ and $k$, we find
\begin{equation}
S_R=\bigl[\text{initial row}\leftrightarrow\text{final row in }\Gt\bigr]\in\{0,1\}.
\end{equation}
The same connection event gives $P_\tau$ from the MMIS (see \cref{sm:info}), which identifies $S_R$ with $P_\tau$.

\emph{Unrecorded locations.} 
The exact mapping requires that the locations of the dephasing events be recorded. Each bond weight is then a projector, so that a bond either fixes the relation between the fields across it or leaves them free. The sign arguments of \cref{sm:sign} treat every bond of a loop or of a cluster as such a hard constraint. If the locations were not recorded, the observer would have to average each bond over whether a dephasing occurred on it. A temporal bond without $X$ measurement and a spatial bond without $ZZ$ measurement would then carry
\begin{equation}
W_{ac}=\tfrac14(1+v)\bigl[1+(1-q)\,u\bigr],
\qquad
W_{ac}=1-\tfrac q2+\tfrac q2\,v ,
\end{equation}
respectively. These weights are no longer projectors. They couple $\sigma$ in time and $\tau$ in space with finite strength, so that a frustrated record acquires a finite Born weight. The doubled state then maps onto a disordered statistical model, of random-bond Ising type in suitable limits~\cite{lee_decoding_2022,hauser_information_2025}, rather than onto percolation. On a processor, errors at unknown locations play a similar role. For a fixed decoder, \cref{sm:exp-err} shows that independent Pauli faults and readout flips then attenuate the mean of its single-shot estimator by an exact factor.

\section{Protocol and algorithm}
\label{sm:protocol}

\emph{Reference qubits.} 
A reference qubit $R$ starts in $|0\rangle$ and is coupled to site $j$ by a controlled-NOT gate from $j$ to $R$, which copies the value of $Z_j$ into $R$. In the doubled representation the gate acts on both the ket copy, labeled $F$, and the bra copy, labeled $B$, so that
\begin{equation}
(s_{R},s'_{R})=(s_j,s'_j)\quad\Longrightarrow\quad(\sigma_R,\tau_R)=(\sigma_j,\tau_j),
\end{equation}
and $R$ carries both fields of site $j$ at that time. In the percolation picture $R$ is a new vertex attached by a single bond to the vertex of $j$, in both graphs, and all other bonds stay unchanged. No later operation acts on $R$, so $R$ keeps this copy until the final time. Two references coupled to the same site at the same time are connected through that site, which gives $A_\alpha(0)=1$ for every initial state. For the decoded $Z$ statistics, replacing the coupling and $Z$ readout of $R$ by a $Z$ measurement of site $j$ removes the extra qubit (\cref{sm:exp-est}).

\emph{Delayed readout.} Physically, each $\kappa$ is a separate run, in which $R_1$ is coupled to site $i$ at time $t$ and the circuit continues to $T=(1+\kappa)t$. In the simulation the trajectory is copied at time $t$ and continued along separate branches for the different $\kappa$, each with its own references, which gives the same statistics. For $\kappa=0$ the site itself can serve as $R_1$. By \cref{conn},
\begin{equation}
A_\alpha(t;\kappa)=\Pr\bigl[(i,0)\leftrightarrow(i,t)\ \text{in a slab of height }(1+\kappa)t\bigr],
\end{equation}
which cannot decrease with $\kappa$ in any trajectory, since added rows can only add paths. The equal-time correlator $C_\alpha(r,t;\kappa>0)$ on the insertion slice uses further references coupled at time $t$ to the sites $i+r$ and is evaluated at $T$. We use $\kappa\in\{1,2\}$, and read effective exponents only where $(1+\kappa)t\le L/8$ (\cref{sm:numerics}).

\emph{Doubled-state algorithm.} The algorithm uses the unnormalized state $\tilde\rho_m$ and evaluates physical expectation values in $\rho_m=\tilde\rho_m/\Tr\tilde\rho_m$. For a stabilizer state $\tilde\rho_m$, the doubled vector $\dket{\tilde\rho_m}$ is itself a pure stabilizer state, on $2L$ qubits for the system and two for each reference. The simulation follows this pure state along each trajectory and evolves its stabilizer group, which gives the Born probabilities exactly. A dephasing of $O$ applies the projector $\tfrac12(1+O_FO_B)$, with $O_{F,B}$ acting on the ket and bra copies,
and leaves $\Tr\tilde\rho_m$ unchanged. With $\mathcal S_m$ the stabilizer group of $\dket{\tilde\rho_m}$, a measurement of $O$ has the outcome probabilities
\begin{equation}
\Pr(\mu)=\frac{\Tr\Pi_{O,\mu}\tilde\rho_m}{\Tr\tilde\rho_m}=
\begin{cases}\tfrac12(1+\mu\epsilon), & \epsilon O_F\in\mathcal S_m,\ \epsilon=\pm1,\\
\tfrac12, & \pm O_F\notin\mathcal S_m,\end{cases}
\end{equation}
and we apply $\Pi_{O,\mu}\otimes\Pi_{O,\mu}$ with the drawn $\mu$ on both copies. The summands of \cref{diag} are
\begin{equation}
\left(\frac{\Tr\tilde\rho_mO}{\Tr\tilde\rho_m}\right)^2=\bigl[\pm O_F\in\mathcal S_m\bigr],\qquad
\frac{\Tr\tilde\rho_mO\tilde\rho_mO}{\Tr\tilde\rho_m^2}=\bigl[O_FO_B\in\mathcal S_m\bigr].
\end{equation}
The initial stabilizer groups are
\begin{equation}
\mathcal S_{\rm PDS}=\langle X_{j,F},X_{j,B}\rangle,\qquad
\mathcal S_{\rm GHZ}=\langle Z_{j,F}Z_{j+1,F},\,Z_{j,B}Z_{j+1,B},\,\bar X_F,\,\bar X_B\rangle,
\end{equation}
with $j=1,\dots,L$. The MMIS is the state reached from the PDS by complete dephasing of every $Z_jZ_{j+1}$. Following Ref.~\cite{weinstein_efficient_2025}, we prepare it by placing the ket copy in the GHZ state and the bra copy in $|{+}\rangle^{\otimes L}$ and applying a controlled-NOT gate from $B_j$ to $F_j$ on every site. Each configuration $s'$ of the bra copy then flips the ket spins at the sites with $s'_j=-1$, which turns $|{\Uparrow}\rangle+|{\Downarrow}\rangle$ into $|s'\rangle+|-s'\rangle$. The result is $\dket{\rho}\propto\sum_{s'}(|s'\rangle+|-s'\rangle)|s'\rangle$, the doubled form of $\rho(s,s')=2^{-L}(\delta_{s,s'}+\delta_{s,-s'})$.

\section{Scaling details}
\label{sm:scaling}
At the transition ($g=0$) and for $1\ll t\ll L^z$, every quantity of \cref{conn} is the probability of a connection in a scale-invariant spacetime slab. Its scaling then follows from the scaling dimensions of the endpoints, which depend on whether an endpoint lies on the initial row, on the final row or inside the slab. This section treats one order parameter at a time and drops the label $\alpha$. \Crefrange{sm:auto}{sm:bulk} treat the autocorrelation, the equal-time correlator and an insertion inside the bulk. \Cref{sm:modelA,sm:aging} compare with relaxational dynamics and consider a waiting time.

Under weak measurements the trajectories are no longer stabilizer states, and the final time may then impose different boundary conditions on the EA and R\'enyi-2 correlators~\cite{hauser_information_2025}. As long as $A_\alpha$ and $C_\alpha$ are measured with the same diagnostic on the same insertion slice, they should share the same $\xobs$, which \cref{master} then still removes. We have not tested this case numerically.

\subsection{Autocorrelation}
\label{sm:auto}

Consider the slab of height $t$, infinite in space, with an endpoint of dimension $x_1$ on the initial-time boundary and one of dimension $x_2$ on the final-time boundary at the same position. Under a rescaling of lengths by $\ell$ and of times by $\ell^z$,
\begin{equation}
A(t)=\ell^{-(x_1+x_2)}A(\ell^{-z}t),
\label{smrescale}
\end{equation}
and, since $\ell$ is arbitrary, the choice $\ell=t^{1/z}$ gives $A\propto t^{-(x_1+x_2)/z}$. A wired boundary, or a row summed over as a whole, is connected to every path that reaches it and carries no dimension, so that
\begin{equation}
A\big|_{\rm wired}\propto t^{-x_2/z},\qquad P\propto t^{-x_1/z},
\label{smwir}
\end{equation}
both one-point functions of a single endpoint.

\subsection{Equal-time correlator}
\label{sm:equal}

Two points on the final row at distance $r$ obey
\begin{equation}
C(r,t)=r^{-2x_s}\,\Phi\bigl(r/t^{1/z},\,r/L\bigr),
\end{equation}
with $\Phi$ a scaling function that depends on the initial boundary condition. For $r\ll t^{1/z}$ the operator product expansion fuses the pair into single operators at its midpoint. Their expectation values feel the initial boundary only through corrections in powers of $r/t^{1/z}$, and $\Phi$ tends to a constant independent of the initial boundary. For a wired initial row this can be bounded directly. A connection of the two points $a$, $c$ that exists only through the wired row requires two disjoint paths, one from each point to the row, and by the van den Berg--Kesten inequality~\cite{van_den_berg_inequalities_1985},
\begin{equation}
\Pr\bigl[a\leftrightarrow c\ \text{only via the row}\bigr]
\le\Pr[a\leftrightarrow\text{row}]\,\Pr[c\leftrightarrow\text{row}]\propto t^{-2x_s/z},
\end{equation}
which is negligible against $r^{-2x_s}$ for $r\ll t^{1/z}$. For $r\gg t^{1/z}$, the two paths become independent and the bound is reached,
\begin{equation}
C(r,t)\to\begin{cases}0, & \text{free initial row},\\ \simeq A^2, & \text{wired initial row},\end{cases}
\end{equation}
with $A\big|_{\text{wired}}$ of \cref{smwir}, so that from a wired row $Cr^\eta$ rises with $r$ instead of falling. For $t\gg L^z$ the influence of the initial boundary vanishes, and
\begin{equation}
C(r,L)\propto\Bigl[\frac{L}{\pi}\sin\frac{\pi r}{L}\Bigr]^{-2x_s},
\label{smchord}
\end{equation}
the chord-length form~\cite{cardy_conformal_1984,calabrese_entanglement_2009}, which
carries the final-time dimension but none from the initial state (\cref{sm:z}).

\subsection{Insertion inside the bulk}
\label{sm:bulk}

If the later endpoint sits at height $t$ in a slab of height $(1+\kappa)t$, the lengths $t$ and $\kappa t$ scale together, and the endpoint is a bulk point for every fixed $\kappa>0$. Scale invariance gives
\begin{equation}
A(t;\kappa)=t^{-(x_1+x_b)/z}\,H(\kappa),
\end{equation}
with $H$ a scaling function. Two points on the insertion slice, a time $\Delta=\kappa t$ below the final row, obey
\begin{equation}
C(r,t,\Delta)=r^{-2x_b}\,\Phi\bigl(r/t^{1/z},\,r/\Delta^{1/z}\bigr),
\end{equation}
which decays as $r^{-2x_b}$ for $r\ll t^{1/z},\Delta^{1/z}$ and crosses over to $\Delta^{2(x_s-x_b)/z}r^{-2x_s}$ for $\Delta^{1/z}\ll r\ll t^{1/z}$. This intermediate regime exists only for $\kappa\ll1$. For the $\kappa=1,2$ used here the slice correlator decays as $r^{-2x_b}$ throughout $r\ll t^{1/z}$. The correlator on the final row still decays as $r^{-2x_s}$, since that row is a boundary at every $\kappa$. Each geometry therefore needs the correlator of its own slice. From a wired initial
row, $x_1=0$, the temporal exponent equals half the spatial one on either slice, $\lambda=\eta/2$, as the MMIS shows for $\tau$ in \cref{slice}(b). The limits $\kappa\to0$ and $t\to\infty$ do not commute, and the whole slab must satisfy $(1+\kappa)t\ll L^z$.

\subsection{Relaxational dynamics}
\label{sm:modelA}
In model A the observables are linear in the probability distribution, and its conservation makes later times irrelevant. The observation time is therefore a bulk point, $\xobs=x_b$, and the equal-time correlator at
that time decays as $r^{-2x_b}$ for $r\ll t^{1/z}$. A small initial magnetization $m_0$ couples to the order-parameter field on the initial slice. With $x_0$ the scaling dimension of $m_0$~\cite{janssen_new_1989,zheng_monte_1998}, that field has dimension $\xini=d-x_0$, with $d$ the spatial dimension, so that the autocorrelation from a disordered state decays as
\begin{equation}
A\propto t^{-(d-x_0+x_b)/z},
\end{equation}
and \cref{master} reads $\lambda=d-x_0+x_b$. A small initial magnetization gives $M(t)=m_0\sum_x A(x,t) \propto m_0t^{\theta'}$, where $A(x,t)$ correlates the order parameter at positions separated by $x$ and at times $0$ and $t$, and $\theta'=(d-\lambda)/z$. Model A thus gives $\theta'=(x_0-x_b)/z$, the initial-slip exponent of the magnetization~\cite{janssen_new_1989,zheng_monte_1998}, and $\lambda=d-z\theta'$ is the known relation~\cite{janssen_phase_1992,calabrese_ageing_2005} between it and the autocorrelation exponent of Ref.~\cite{huse_remanent_1989}. As in Ref.~\cite{huse_remanent_1989}, $\lambda$ is divided by $z$ in the decay of $A$, which makes it a sum of scaling dimensions. Refs.~\cite{zheng_monte_1998,yu_nonequilibrium_2026}
quote the exponent of $t$ itself, $\lambda/z=d/z-\theta'$, which equals $\lambda$ here since $z=1$. The exponent of the response function, also called the initial-slip exponent and denoted $\theta$, differs from $\theta'$~\cite{calabrese_ageing_2005}.

In the circuit the observed slice is the final row. From a free initial row the spatially integrated connectivity, the mean number of final-row sites connected to an initial site, therefore grows as
\begin{equation}
\sum_xA(x,t)\propto t^{(d-2x_s)/z}=t^{1/3},
\end{equation}
not as $t^{(d-x_s-x_b)/z}=t^{9/16}$, the value for a bulk observation point. The connection reaches the final row with probability $\propto t^{-x_s/z}$ and then covers $\propto t^{(d-x_s)/z}$ of its sites. Identifying this growth exponent with the initial slip of a physical magnetization would require a protocol with a small initial bias and a specified average over records, which we do not construct.

\subsection{Waiting time}
\label{sm:aging}

Coupling $R_0$ at a waiting time $t_w$ instead of $t=0$ places the earlier endpoint a time $t_w$ above the initial row. The autocorrelation between the times $t_w$ and $t_w+t$ then obeys
\begin{equation}
A(t_w+t,\,t_w)=t^{-(x_b+x_s)/z}\Psi_{\xini}(t/t_w),
\end{equation}
with $\Psi$ a scaling function that depends on the initial boundary condition. For $t\ll t_w$ the earlier endpoint is a bulk point, $\Psi$
tends to a constant independent of the preparation, and the leading scaling no longer depends on the initial state. For $t\gg t_w$ the bulk-to-boundary expansion gives $\Psi\propto(t/t_w)^{(x_b-\xini)/z}$, so that
\begin{equation}
A\propto t_w^{(\xini-x_b)/z}\,t^{-(\xini+x_s)/z},
\end{equation}
and the scaling set by the initial boundary condition is recovered, as in critical aging~\cite{calabrese_ageing_2005}. These scaling predictions apply for $1\ll t_w$ and $t_w+t\ll L^z$. We have not tested them numerically.

\section{Away from the transition}
\label{sm:off}

\emph{Scaling form.} Away from the transition a rescaling by $\ell$, as in \cref{smrescale}, also multiplies $g$ by $\ell^{1/\nu}$. With the finite size kept,
\begin{equation}
\begin{aligned}
A_\alpha(t,g,L)&=\ell^{-\lambda_\alpha}A_\alpha\bigl(\ell^{-z}t,\,\ell^{1/\nu}g,\,L/\ell\bigr),\\
C_\alpha(r,t,g,L)&=\ell^{-\eta_\alpha}C_\alpha\bigl(r/\ell,\,\ell^{-z}t,\,\ell^{1/\nu}g,\,L/\ell\bigr),
\end{aligned}
\label{eq:smscaling}
\end{equation}
and the choice $\ell=t^{1/z}$ gives \cref{scalingA} with $\mathcal{F}_\alpha (g t^{1/\nu z},t/L^z)=A_\alpha (1,g t^{1/\nu z},L/t^{1/z})$. On the disordered side
\begin{equation}
\ln A_\alpha=-c_0\,t/\xi_t+O(\ln t),\qquad \xi_t\propto|g|^{-\nu z},
\label{eq:smdecay}
\end{equation}
with a constant $c_0$. On the ordered side, with the thermodynamic limit taken first, $A_\alpha$ saturates. A plateau independent of $t$ requires $\mathcal F_\alpha\propto(|g|\,t^{1/\nu
z})^{\nu\lambda_\alpha}$ at large argument, so that
\begin{equation}
\lim_{t\to\infty}\lim_{L\to\infty}A_\alpha(t,g,L)\propto|g|^{\nu\lambda_\alpha}.
\label{eq:smplateau}
\end{equation}
In the percolation picture this plateau is the product of the probabilities that the two endpoints belong to the infinite cluster, each
$\propto|g|^{\nu x}$ with $x$ the dimension of its endpoint ($x=0$ on a wired row). The order of limits matters. At finite $L$ and $p>0$, every time step measures all $X_i$ with probability $p_x^L$, which removes every temporal bond of both graphs in that step, so that
\begin{equation}
A_\alpha(t,L)\le(1-p_x^L)^t\xrightarrow{t\to\infty}0.
\label{eq:smmeta}
\end{equation}
At finite size the plateau is therefore metastable.
\Cref{hide}(b) collapses $A_\sigma$ from the PDS on both sides of $p_c^\sigma$ with the exact $\lambda_\sigma$, $\nu$ and $z$. Another preparation has its own $\lambda_\sigma$ and $\mathcal F_\sigma$ but the same scaling variable.

\emph{Survival probabilities.} The survival probabilities depend on how the initial state stores the bit. An initial state disordered in an order parameter holds the bit at site $i$ alone. Its initial row is free, and $P_\alpha$ behaves like $A_\alpha$,
\begin{equation}
P_\alpha^{\rm free}\simeq
\begin{cases}e^{-c\,t/\xi_t}, & \alpha\ \text{disordered},\\
\Pr[(i,0)\in\text{infinite cluster}]\propto|g|^{\nu x_s}, & \alpha\ \text{ordered},\end{cases}
\label{eq:smPfree}
\end{equation}
with the same order of limits. For a wired row, $1-P_{\alpha}^{\text{wired}}$ is the probability of a winding dual path, and from the MMIS $P_{\tau}=S_R$ (see \cref{sm:info}). With $\xi\propto |g|^{-\nu}$,
\begin{equation}
P_\tau^{\rm wired}\simeq
\begin{cases}\min\bigl[1,(L/\xi)\,e^{-t/\xi_t}\bigr], & \text{unbroken phase},\\
1-O\bigl(t\,e^{-c'L/\xi}\bigr), & \text{SW-SSB phase},\end{cases}
\label{eq:smPwired}
\end{equation}
with a constant $c'$. In the unbroken phase $\Gt$ consists of finite clusters of size $\xi$. A connection from the initial to the final row then needs one of about $L/\xi$ independent columns to stay connected over the height $t$, which has probability of order $(L/\xi)e^{-t/\xi_t}$. In the SW-SSB phase $\Gt$ percolates, and a winding dual path that cuts it must cross the whole system. Its probability per time step is therefore exponentially small in $L/\xi$.

\emph{Correlation time.} To avoid assuming a form for the crossover, we define the correlation time $\xi_t^\sigma$ near $p_c^\sigma$ through the ratio to the critical curve with the same initial state and size,
\begin{equation}
\mathcal R(t,g)=\frac{A_\sigma(t,g)}{A_\sigma(t,0)}=\frac{\mathcal F_\sigma(g\,t^{1/\nu z},\,t/L^z)}{\mathcal F_\sigma(0,\,t/L^z)},
\label{eq:smratio}
\end{equation}
in which the power-law prefactor cancels. The correlation time $\xi_t^\sigma$ is the time at which $\mathcal R=1/e$. For $t\ll L^z$ the second argument drops out. On the disordered side $\mathcal R$ then falls below any fixed level and depends on $t$ only through $t/\xi_t$, so that
\begin{equation}
\xi_t^\sigma=c_{\xini}\,g^{-\nu z},
\label{smtauh}
\end{equation}
with an amplitude $c_{\xini}$ set by the preparation and an exponent set by the bulk alone. \Cref{smtauh} requires $\xi_{t}\ll L^{z}$, which we check in \cref{sm:numerics}.

\emph{Information erasure.} Beyond $p_c^\tau$, $A_\tau$ decays as well, on a time $\xi_{t}^\tau$ defined as $\xi_{t}^\sigma$ with $A_\tau$ in place of $A_\sigma$. By the Kramers--Wannier duality of \cref{sm:cri}, $\Gt$ at $p_c^\tau+g$ is distributed as the planar dual of $\Gs$ at $p_c^\sigma-g$, so the correlation times of the two graphs diverge with the same exponent and
$\xi_{t}^\tau\propto g^{-\nu z}$. We have not simulated this regime. Together with \cref{hide} and the transition of the reference entropy in \cref{sm:z}, this completes the picture of the main text for a bit held at one site, as in the PDS. At late times
\begin{equation}
\bigl(P_\sigma,\,P_\tau\bigr)\to
\begin{cases}(>0,\ >0), & \text{total SSB: decodable},\\
(0,\ >0), & \text{SW-SSB: hidden after }\xi_{t}^\sigma,\\
(0,\ 0), & \text{unbroken: erased after }\xi_{t}^\tau.\end{cases}
\label{eq:smfates}
\end{equation}
A bit shared by all sites is lost only through a winding dual path (see \cref{sm:info}).

\section{Numerical procedures and errors}
\label{sm:numerics}

\emph{Simulation.} The doubled state is evolved exactly with the algorithm of \cref{sm:protocol}. Each trajectory is prepared afresh, $A_\alpha(0)=1$ is checked before the first layer, and for $\sigma$ the square is taken trajectory by trajectory before averaging. This simulation does not presuppose the percolation mapping, so that its agreement with the predictions also tests the mapping at sizes far beyond the exact checks of \cref{sm:verify}. Where only connection probabilities are needed, as for the other working points of \cref{sm:cri}, the survival probabilities and the error analysis in this section, sampling the graphs of \cref{bonds} directly gives the same averages for larger $L$.

\emph{Effective exponents.} The local slopes $\lambda_\alpha(t)=-z\,d\ln A_\alpha/d\ln t$ of \cref{vector} are evaluated by least squares over a sliding window of half a decade. The lower end of their flat part is set by lattice effects and the upper end by finite size, identified as the time at which the data for successive sizes in \cref{vector} separate, $t\approx L/8$ from a free initial row.

The effective exponents of \cref{slice}(b), at $L=512$ alone, are read only up to the same
height of the slab, $(1+\kappa)t\le L/8$. The points shown beyond it, up to $t=L/8$, fall
below $1/3$ for the PDS by at most $0.02$. Within this window $\xini^\tau$ of the PDS
at $\kappa=1,2$ decreases from $0.43$ at $t\approx4$ to $0.33$, with errors of about $0.005$,
without a plateau. Its value $0.33(1)$ in \cref{results} refers to $\kappa=0$. Beyond a lattice-scale overshoot at
$t\lesssim5$, the approach is set by the free boundaries alone, as the four combinations of
\cref{slice}(b) show. With a wired initial row and the later insertion in the bulk (MMIS,
$\kappa>0$), $\lambda$ is flat from $t\approx4$. A free final row makes it approach from below
(MMIS, $\kappa=0$), a free initial row from above (PDS, $\kappa>0$), and with both
(PDS, $\kappa=0$) the final row prevails. On the final row this is the boundary
correction that also lowers the measured $\eta$ by less than $1\%$. At $L=512$ both limits are
reached before finite size sets in. From a free initial row $\lambda$ reads $0.66(1)$ from
$t\approx10$ on, and from a wired one it approaches $1/3$ from below and reads $0.33(1)$ from
$t\approx45$ on.

\begin{figure}[htb]
\centering
\includegraphics[width=4.9in]{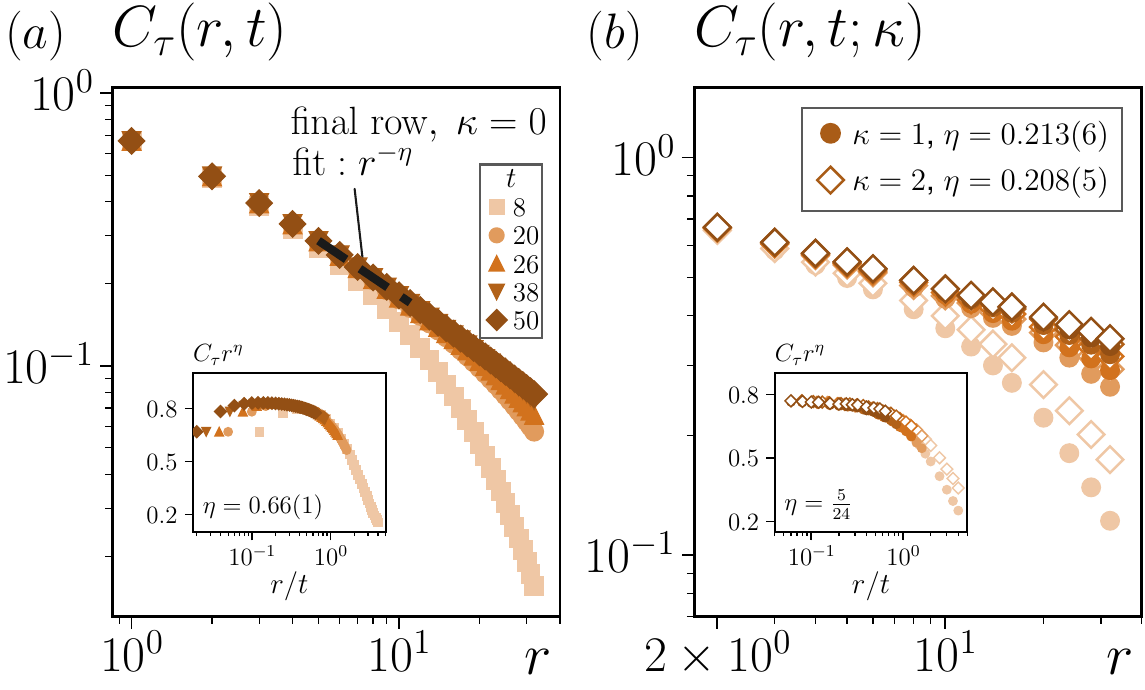}
\caption{\textbf{Equal-time correlators decay with the boundary exponent on the final row and with the bulk exponent on the insertion slice.} Data for $\tau$ at $p_c^\tau$ and $L=512$ from the PDS, at $t=8,\dots,50$ (pale to deep). \textbf{(a)}~On the final row ($\kappa=0$), the fit $r^{-\eta}$ at $t=50$ (dashed) gives $\eta=0.66(1)$, consistent with $2x_s=2/3$. Inset: $C_\tau r^\eta$ against $r/t$ with $\eta=0.66$. \textbf{(b)}~On the insertion slice for $\kappa=1$ (filled circles) and $\kappa=2$ (open diamonds), the correlator decays more slowly, with $\eta$ extrapolated in $1/t$ (\cref{S1}) and consistent with $2x_b=5/24$. Inset: $C_\tau r^{\eta}$ against $r/t$ for $r\ge3$, with the exact $\eta=5/24$. In both insets the data for different $t$ collapse, so that the initial row enters only through $r/t$. Error bars are one standard error.}
\label{Sslice}
\end{figure}

\begin{figure}[htb]
\centering
\includegraphics[width=1.95in]{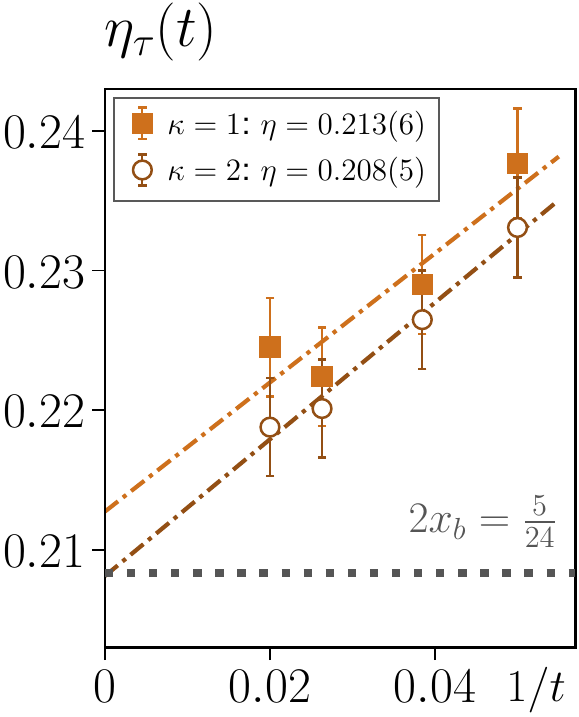}
\caption{\textbf{Extrapolated in $1/t$, the equal-time exponent on the insertion slices agrees with $2x_b=5/24$.} Effective exponents $\eta_\tau(t)$, the negative slopes of $\ln C_\tau$ against $\ln r$ over $2\le r\le6$, versus $1/t$ for $\kappa=1$ (filled squares) and $\kappa=2$ (open circles) at $L=512$ and $t=20,26,38,50$. Dash-dotted lines are weighted linear extrapolations to $1/t=0$, and the dotted line marks $2x_b=5/24$. Error bars are one standard error.}
\label{S1}
\end{figure}

\emph{Equal-time exponents.} The two slices are read differently because their leading
corrections differ. On the final row, a free boundary is defined only to within a lattice
spacing, which adds a relative correction of order $1/r$, while the initial row enters only at
higher order in $r/t$. The product $C\,r^{2x_s}$ is therefore flat at intermediate distances, and $\eta$
is the negative slope of a straight-line fit of $\ln C$ against $\ln r$ at $t=50$ over $5\le r\le11$,
$\eta_\tau=0.664(2)$. The two corrections pull in opposite directions, and moving the window or
the final time changes this value by up to about $0.01$, the error quoted in the main text. A
pair on an insertion slice has no boundary correction but feels both boundaries, at
distances $t$ and $\kappa t$, so that $C\,r^{2x_b}$ decreases from the smallest $r$ on,
nearly linearly in $r/t$, and has no flat part. On the insertion slices, the negative slope over $2\le r\le6$ defines $\eta_\tau(t)$ at each $t\ge20$. Extrapolating it linearly in $1/t$ (\cref{S1}) gives
$\eta=0.213(6)$ and $0.208(5)$ for $\kappa=1,2$, against $2x_b=5/24$. This extrapolation is
empirical. It extrapolates away the dependence on the slab height but does not eliminate lattice corrections at fixed small $r$. A fit at a single time would overestimate $\eta$ by about $0.01$ at $t=50$. We take this shift as an estimate of the systematic uncertainty of the extrapolated values. It is not included in the quoted statistical errors. All fits use $L=512$.

\begin{figure}[htb]
\centering
\includegraphics[width=4.9in]{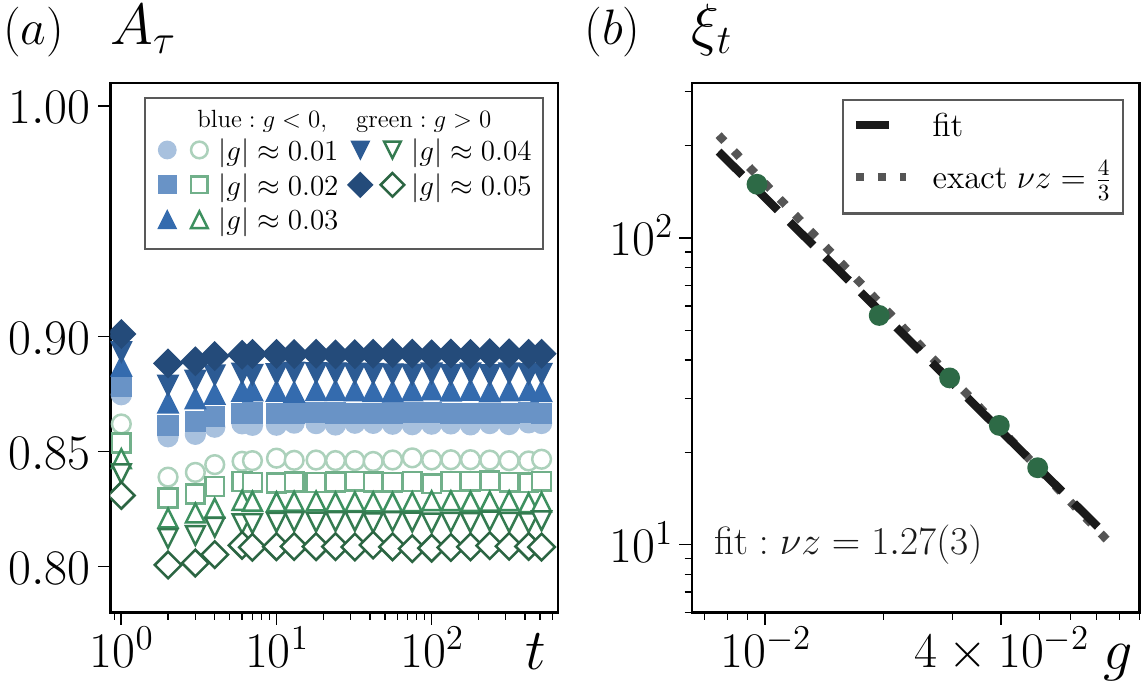}
\caption{\textbf{Near $p_c^\sigma$, $A_\tau$ saturates while the correlation time of $A_\sigma$ grows as $g^{-\nu z}$.} Data from the PDS at $L=512$, with the symbols of \cref{hide}. \textbf{(a)}~$A_\tau$ saturates within a few time steps on both sides of $p_c^\sigma$, since $\tau$ is ordered throughout. \textbf{(b)}~The correlation time $\xi_t$ against $g>0$ follows a weighted power-law fit with $\nu z=1.27(3)$ (dashed), compared with the exact $4/3$ (dotted, normalized at the largest $g$). Error bars are one standard error.}
\label{Sxi}
\end{figure}

\emph{Correlation time.} The crossing of $\mathcal R$ with a level is located by a straight line of
$\ln\mathcal R$ against $\ln t$ through the sampled times within $15\%$ of it, starting from the
first time below the level. The window is fixed in $\ln t$ because successive sampled times
differ by only a few percent, so that a few of them would leave the local slope to the noise
of $\mathcal R$. Only crossings later than about $8$ steps are used, below which the
discreteness of time biases the interpolation. A weighted straight-line fit of
$\ln\xi_t^\sigma$ against $\ln g$ at $L=512$ gives $\nu z=1.27(3)$ for $g\le0.05$, the value in
\cref{Sxi}(b), and the levels $\mathcal R=0.5$ and $0.2$ give $1.28(3)$ and $1.29(2)$, all
about two standard errors below the exact $4/3$. The deviation is not concentrated at the
smallest $g$, where $\xi_t^\sigma\approx150$ exceeds $L/8$. The local slopes between successive
$g$ are $1.37(6)$, $1.13(9)$, $1.22(15)$ and $1.41(17)$ at the level $1/e$, and the smallest pair
gives $1.32$ to $1.37$ at the three levels. At $g=0.0095$ the size $L=256$ gives
$\xi_t^\sigma=158(7)$, within one standard error of $L=512$, and at every $g$ the two sizes agree
within $2.3$ standard errors.

One candidate cause is a correction analytic in $g$, since the
bond probabilities, and with them the anisotropy that converts the correlation length into a
time, change linearly with $g$. With the exact exponent and a coefficient $b$,
\begin{equation}
\xi_{t}^\sigma \,g^{4/3}=c_{\xini}(1+b\,g)
\label{smtauhcorr}
\end{equation}
fits every level, each with its own amplitude, with $\chi^2\le4.3$ for three degrees of
freedom and $b$ between $1.6$ and $2.9$, a rise of $6$ to $12\%$ across the range. The present
data do not isolate the origin of the deviation. Corrections to scaling and the fitting window
could contribute as well. The collapse in \cref{hide}(b), which
requires the same exponent at every level, holds to this accuracy.

\begin{figure}[htb]
\centering
\includegraphics[width=4.9in]{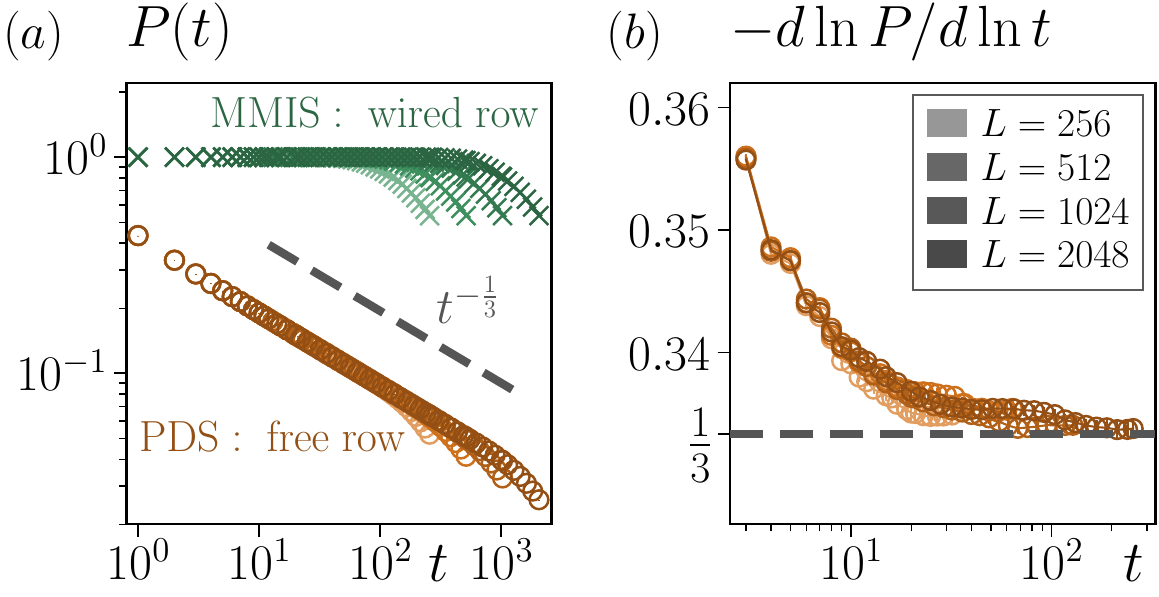}
\caption{\textbf{The stored bit decays as $t^{-x_s/z}$ from a free initial row and survives from a wired one.} Survival probability $P$ at the critical point, from direct sampling of the percolation graphs for $L=256$, $512$, $1024$ and $2048$ (pale to deep). \textbf{(a)}~$P(t)$ from a free initial row (PDS, orange circles), which also gives $P_\sigma$ of the MMIS, and from a wired one (MMIS, green crosses), which also gives both $P_\alpha$ of the GHZ state. The dashed line is $t^{-1/3}$. \textbf{(b)}~Local exponent of the free row over half a decade, up to $t=L/8$, approaching $x_s/z=1/3$ (dashed). Error bars are one standard error.}
\label{SP}
\end{figure}

\emph{Survival probabilities at the critical point.} Since $P_\alpha$ is a connection probability, we sampled the graphs of \cref{bonds} at the critical bond probabilities of \cref{B1} directly (\cref{SP}). We used rings of $L=256$ to $2048$ with $2^{18}$ trajectories per size in $64$ independent batches and averaged the free initial row over its sites. A free initial row describes $P_\sigma$ and $P_\tau$ of the PDS and $P_\sigma$ of the MMIS. Its local exponent $-d\ln P/d\ln t$ decreases towards $x_s/z=1/3$ and reads $0.3338(7)$ at $t=256$ for $L=2048$, while straight-line fits of $\ln P$ against $\ln t$ over $20\le t\le L/8$ and $36\le t\le L/8$ give $0.334$ to $0.336$ for $L=512$ to $2048$, with batch errors below $0.001$. We quote $0.334(2)$, whose error covers this dependence on the window. A wired initial row describes $P_\tau$ of the MMIS and both survival probabilities of the GHZ state. There $P$ exceeds $0.998$ up to $t=L/8$ for every $L$ and then decays as a function of $t/L$ alone, to $0.964$, $0.808$ and $0.536$ at $t/L=1/4$, $1/2$ and $1$. For the MMIS this is the reference entropy $S_R$ of \cref{sm:z}, and the two agree within $2.2$ standard errors at all $39$ common points of $L=256$ and $512$.

\begin{table}[htb]
\caption{Exponents not listed in Table~I of the main text, predicted by the exact mapping and obtained from the simulations of this section. The equal-time exponents on the insertion slices are for $\tau$ at $p_c^\tau$ from the PDS at $L=512$ (\cref{Sslice,S1}). The errors quoted for these slice exponents are statistical only, with an additional systematic uncertainty of about $0.01$. On these slices the MMIS gives $\xini^\tau=0$, as on the final row [\cref{slice}(b)]. The decay exponents of $P_\alpha$, equal to $\xini^\alpha/z$, come from direct sampling of the percolation graphs at $L=2048$ (\cref{SP}). A free row describes the PDS and $\sigma$ of the MMIS, and a wired row describes $\tau$ of the MMIS and the GHZ state. The correlation-time exponent is for $\sigma$ from the PDS near $p_c^\sigma$ at $L=512$ (\cref{Sxi}).}
\label{SMresults}
\begin{ruledtabular}
\begin{tabular}{llll}
quantity & case & predicted & simulated\\
\hline
$\eta_\tau$ & slice, $\kappa=1,\,2$ & $5/24$ & $0.213(6)$, $0.208(5)$\\
$P_\alpha$ exponent & free row & $1/3$ & $0.334(2)$\\
 & wired row & $0$ & $0.000(1)$\\
$\nu z$ & near $p_c^\sigma$ & $4/3$ & $1.27(3)$\\
\end{tabular}
\end{ruledtabular}
\end{table}

\emph{Errors.} For \crefrange{vector}{hide} and \cref{Sslice,Sxi,S1,S2}, each point averages over $7\times10^5$ trajectories in \cref{vector}
and on the critical curve of \cref{hide}, $3.5\times10^5$ for the PDS in
\cref{slice}, $10^4$ in \cref{S2} and $7\times10^4$ otherwise, with its standard
error. Fits are weighted by the inverse squared errors. The errors of the effective and fitted exponents include the correlations between the times and distances of one run. We obtained them by a jackknife over independent batches, either of the runs themselves or of an ensemble sampled directly on the percolation graphs with the same numbers of trajectories, which by the exact mapping has the distribution of the stabilizer data. The errors in \cref{results} also cover the dependence on the fitting window, as do those of the survival-probability exponents in \cref{SMresults}.

\section{Finite-size regime}
\label{sm:z}

The short-time results hold for $t\ll L^z$ and use no input from later times. At
$t\gtrsim L^z$ the system size enters, and this regime supplies three independent checks: the
critical point, the exponents $\nu$ and $z$ that enter \cref{scalingA}, and the loss
of the initial state from the equal-time correlator.

\emph{Reference entropy.} Following Ref.~\cite{hauser_information_2025}, a reference qubit $R$ is entangled with the strong charge. The joint initial state is $(|0\rangle_R|{\Uparrow}\rangle+|1\rangle_R|{\Downarrow}\rangle)/\sqrt2$, stabilized by $\bar XX_R$, so that $R$ is maximally entangled with $\bar X$. The state of $R$ thus records which charge sector, $\bar X=1$ or $\bar X=-1$, the system occupies. Measurements whose outcomes jointly reveal $\bar X$ also reveal this information, and $R$ then becomes pure. The reference entropy
\begin{equation}
S_R=S(\rho_{R,m}),
\label{smprobes}
\end{equation}
with $S(\cdot)$ the von Neumann entropy in bits, is averaged over trajectories. The entropy $S_R$ vanishes exactly when the record has determined the charge, so that its transition is that of $\tau$~\cite{hauser_information_2025}. It takes the values $0$ or $1$ in each trajectory, and since every operation commutes with $\bar X$, a charge once determined stays determined. Its average is thus the probability that the record has not yet determined the charge. In the percolation picture this is the probability that the wired initial row is connected to the final row in $\Gt$, which is also $P_\tau$ from
the MMIS (see \cref{sm:info,sm:verify}). It vanishes at long times for any finite $L$ and $p>0$, and its scaling form carries no power-law prefactor,
\begin{equation}
S_R(t,L,g)=G\bigl(t/L^{z},\,g\,L^{1/\nu}\bigr),
\label{smUpsilon}
\end{equation}
with $G$ a scaling function.

\begin{figure*}[htb]
\centering
\includegraphics[width=7in]{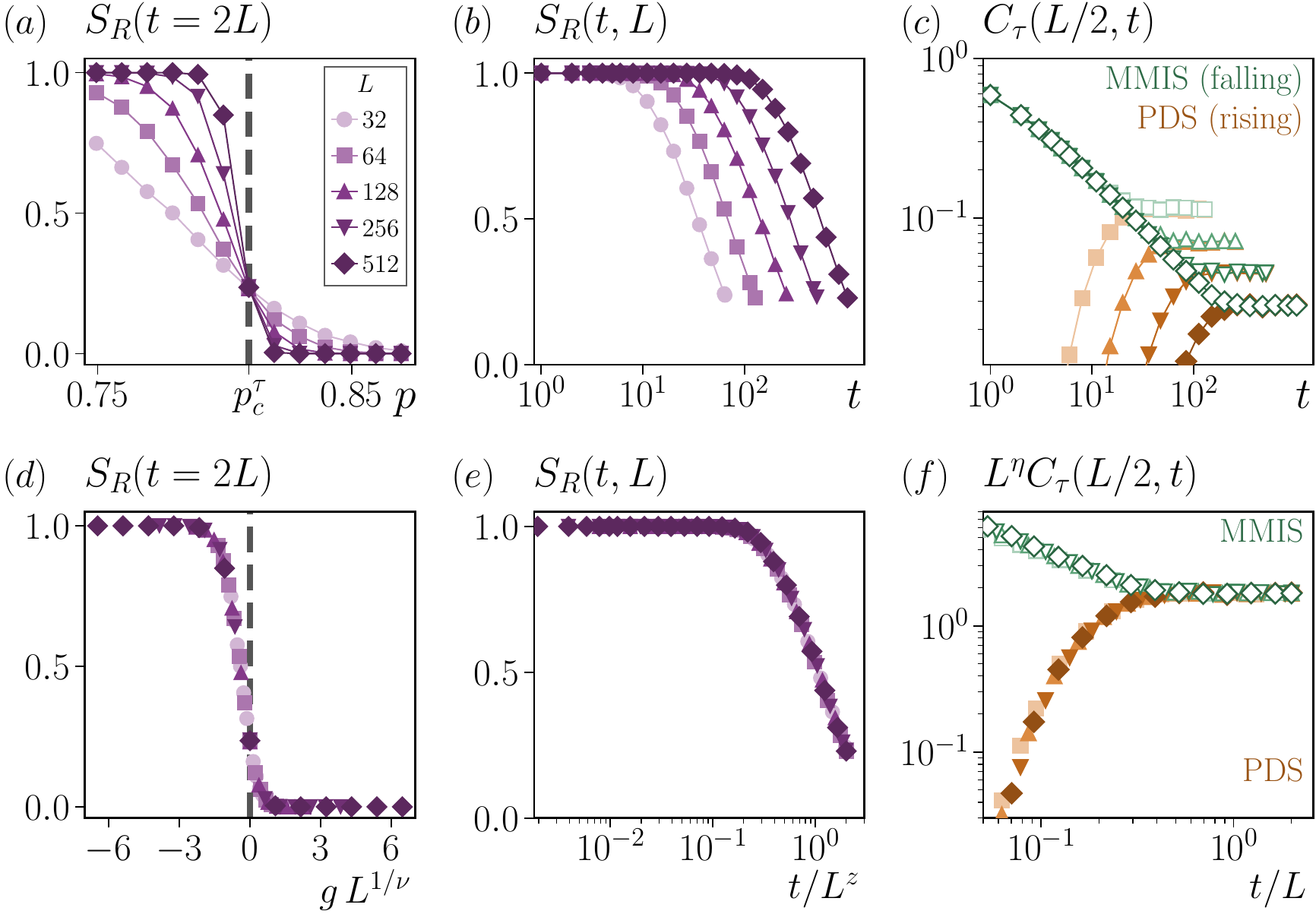}
\caption{\textbf{For $t\sim L^z$ the reference entropy crosses at $p_c^\tau$ and the initial state is forgotten.} The top row shows the data and the bottom row their collapse, with shades from pale (small $L$) to deep (large $L$). \textbf{(a)} Reference entropy at $t=2L$ against $p$ for $L=32,\dots,512$, crossing at $p_c^\tau$. \textbf{(b)} $S_R(t,L)$ at $p_c^\tau$ against $t$. \textbf{(c)} $C_\tau(L/2,t)$ at $p_c^\tau$ from a free (PDS, orange filled symbols, rising) and a wired (MMIS, green open symbols, falling) initial row, $L=64,\dots,512$. \textbf{(d)} The data of (a) against $g\,L^{1/\nu}$, with $g=p-p_c^\tau$ and $\nu=4/3$. \textbf{(e)} The data of (b) against $t/L^{z}$ with $z=1$. \textbf{(f)} The data of (c) as $L^{\eta}C_\tau$ against $t/L$ with $\eta=2/3$. Error bars are one standard error.}
\label{S2}
\end{figure*}

\emph{Critical point and $\nu$.} At the fixed aspect ratio $t/L=2$, which fixes the first argument of $G$ since $z=1$, $S_R$ for $L=32,\dots,512$ crosses at $p_c^\tau$, where it is independent of $L$ within errors, $0.234(2)$ with $\chi^2=1.9$ for four degrees of freedom [\cref{S2}(a)]. This crossing value depends on the aspect ratio and the protocol and is not universal. With the exact $p_c^\tau$ and $\nu=4/3$ the data collapse within errors against $g\,L^{1/\nu}$ without adjustable parameters [\cref{S2}(d)].

\emph{Dynamic exponent.} At $g=0$ the only time scale is $L^z$, and $S_R(t,L)$ for $L=32,\dots,512$ collapses against $t/L^{z}$ with $z=1$ [\cref{S2}(e)]. Since no other exponent enters, this checks $z$ independently of the initial boundary condition and of the final row, and, together with $\nu$, the combination $\nu z$ used in \cref{hide}, without
reference to $\xi_{t}$. These collapses with the known exponents are consistency checks, and they do not exclude small corrections.

\emph{Loss of the initial state.} At the midpoint $r=L/2$ the equal-time correlator from a free (PDS) and a wired (MMIS) initial row approaches the same value from below and from above, respectively, and coincides within errors from $t\approx L/2$ on. Scaled by $L^{\eta}$ with $\eta=2/3$, the data for $L=64,\dots,512$ collapse against $t/L$ [\cref{S2}(f)].
The common value is the chord-length form of \cref{smchord} at $r=L/2$, $\propto L^{-\eta}$ with the same $\eta$ as at short times. The difference between the two comes from connections through the initial row, which become rare once $t\gtrsim L^z$.

\section{Environment-assisted decoding protocol}
\label{sm:exp}

This section supplies the derivations and data behind the decoding protocol that the main text summarizes and \cref{em:proposal} specifies. It shows that every correlator of the ideal circuit has a single-shot estimator without postselection, that independent unrecorded errors attenuate this estimator by an exact factor, and that for bulk errors this factor nearly cancels in the contrast between the MMIS and the PDS. \Cref{sm:exp-est} derives the estimator of \cref{expest} and its shot cost. \Cref{sm:exp-prep} shows that one round of measurements prepares the MMIS. \Cref{sm:exp-err} derives the attenuation of \cref{expvis} and evaluates the visibility collapse and the decoded contrast of \cref{expt}(b,c). \Cref{sm:exp-gate} gives the gate-level compilation behind the error probabilities, the visibility and the shot cost quoted in the main text.

Throughout, $A_\alpha$ and $C_\alpha$ denote the correlators of the ideal circuit, $B_\alpha$ the decoded correlators measured with errors, and an overline the average over shots. Deliberately applied dephasing has known locations and stored labels, whereas hardware faults have unknown realizations and are described by prescribed probabilities. Decoding paths depend on the event pattern and on these probabilities only.

\subsection{Single-shot estimators without postselection}
\label{sm:exp-est}

Each shot combines endpoint readouts with ordinary measurement outcomes and separately stored labels. The event pattern and the initial boundary condition determine $\mathcal G_\alpha$. Both decoders use the record, while only the $\tau$ decoder uses the labels. An $X_i$ measurement is a $Z_i$ measurement between Hadamard gates. A $Z_iZ_{i+1}$ measurement can be performed without an ancilla,
\begin{equation}
\mathrm{CX}_{i\to i+1}\,\Pi_{Z_{i+1},\mu}\,\mathrm{CX}_{i\to i+1}=\Pi_{Z_iZ_{i+1},\mu},
\label{smgadget}
\end{equation}
with $\mathrm{CX}_{i\to i+1}$ the controlled-NOT gate from $i$ to $i+1$ and $\Pi_{O,\mu}$ the projector of \cref{ops}. Alternatively, two controlled-NOT gates map the parity onto an ancilla, which is reusable without reset by taking the product of its successive readouts. To unravel a dephasing of $Z_iZ_{i+1}$, we perform the parity measurement but store its outcome $\mu$ only as a label. To unravel a dephasing of $X_i$, we apply $X_i^{n}$ with a fair random bit $n$ and store $n$ in the same way. Discarding the labels restores the dephasing channels. This construction is equivalent to measuring the purifying environment qubits of \cref{sm:info} right after their coupling.

The reference qubits of \cref{sm:protocol} undergo no further operations after copying $Z_i$. Their readouts can therefore be moved right after the coupling, without changing the joint law of the outcomes, provided that no later operation depends on the results. The coupling and readout can then be replaced by direct $Z_i$ measurements at times $0$ and $t$, with outcomes $s_i(0)$ and $s_i(t)$.

For each occupied bond $\langle a,c\rangle$ of $\Gs$ or $\Gt$, set by the events as in \cref{weight}, the ideal mapping imposes $s_as_c=\mu_{ac}$, with $s_a$ the value of $Z$ at the spacetime point $a$. The sign $\mu_{ac}$ is the parity outcome on a spatial bond, $+1$ on an untouched temporal bond and $(-1)^n$ on a temporal bond with an unraveled $X$ dephasing. The intermediate variables $s_a$ express these constraints and need not be measured. For two endpoints $a_1$ and $a_2$, let $I_\alpha=1$ if $\mathcal G_\alpha$ connects them and $0$ otherwise. For a connected pattern, choose a path $\mathcal P_\alpha$ between them independently of the measurement outcomes and define
\begin{equation}
Y_\alpha=I_\alpha\,\mu_{\mathcal P_\alpha}s_{a_1}s_{a_2},\qquad \mu_{\mathcal P_\alpha}=\prod_{\langle a,c\rangle\in\mathcal P_\alpha}\mu_{ac},
\label{smest}
\end{equation}
with $Y_\alpha=0$ for disconnected patterns. Because multiplying the bond constraints along the path cancels the intermediate signs, $Y_\alpha=I_\alpha$ on every shot of the ideal circuit. Disconnected shots contribute zero and are retained, and no postselection is required. For $a_1=(i,0)$ and $a_2=(i,t)$, \cref{conn} then gives \cref{expest}. The same holds for $C_\alpha$ with $s_i(t)s_{i+r}(t)$, and for an insertion slice with $s_i(0)s_i(t)$ or $s_i(t)s_{i+r}(t)$ measured at the time $t$ of the slice, with paths anywhere in the slab. The decoder must not use the reference or slice outcomes of other sites, which would anchor the clusters they touch like a wired row. \Cref{alg:env-decoding} summarizes the protocol for $A_\alpha(t)$.

\par\medskip\noindent
\begin{minipage}{\linewidth}
\refstepcounter{algorithm}\label{alg:env-decoding}
\hrule\smallskip
\textbf{Algorithm \thealgorithm. Environment-assisted decoding of $A_\alpha(t)$.}
\smallskip\hrule\smallskip
\small
\begin{algorithmic}[1]
\Require Initial state (PDS or MMIS), $L$, circuit parameters, $\alpha\in\{\sigma,\tau\}$, site $i$, time $t$ and number of shots $N$.
\For{$j=1,\ldots,N$}
\State Prepare $|{+}\rangle^{\otimes L}$. For the MMIS, measure one round of $Z_\ell Z_{\ell+1}$ and keep the outcomes as labels.
\State Measure $Z_i$ to obtain $s_i(0)$. Sample a fresh event pattern and run it for $t$ steps:
\State \hspace{1em}Measurements: keep the outcomes in the record.
\State \hspace{1em}$ZZ$ dephasing: measure $Z_\ell Z_{\ell+1}$ and keep the outcome only as a label.
\State \hspace{1em}$X$ dephasing: apply $X_\ell^{n}$ with a fair random bit $n$ and keep $n$ only as a label.
\State Measure $Z_i$ to obtain $s_i(t)$. Construct $\mathcal G_\alpha$ from the event pattern, with the round's bonds only for the MMIS and $\alpha=\tau$.
\State Set $Y_\alpha^{(j)}\gets0$.
\If{$(i,0)$ and $(i,t)$ are connected in $\mathcal G_\alpha$}
\State Choose a path $\mathcal P_\alpha$ from the event pattern and prescribed error probabilities, independently of the outcomes.
\State Read its signs from the record, and for $\alpha=\tau$ also from the labels. Set $Y_\alpha^{(j)}\gets\mu_{\mathcal P_\alpha}s_i(0)s_i(t)$.
\EndIf
\EndFor
\State \Return $N^{-1}\sum_{j=1}^{N}Y_\alpha^{(j)}$, keeping all shots. It estimates $A_\alpha$ in the ideal circuit and $B_\alpha$ with errors.
\end{algorithmic}
\smallskip\hrule
\end{minipage}
\par\medskip

Since $Y_\alpha^2=I_\alpha$, estimating $A_\alpha>0$ with a relative root-mean-square error $\delta$ requires
\begin{equation}
N\simeq\frac{1-A_\alpha}{\delta^2A_\alpha}\propto t^{\lambda_\alpha/z}
\label{smcost}
\end{equation}
shots, each with a fresh event pattern, where the scaling holds at criticality for small $A_\alpha$. Under the error model of \cref{sm:exp-err} the mean becomes $B_\alpha=A_\alpha V_\alpha$, and estimating $B_\alpha$ to the same relative precision requires $N\simeq\delta^{-2}[1/(A_\alpha V_\alpha^2)-1]$. Recovering $A_\alpha$ from this mean additionally requires knowledge of $V_\alpha$, and uncertainty in its calibration is not included in this shot count. The expressions for $C_\alpha$ follow by replacing $A_\alpha$ with $C_\alpha$.

\subsection{Mixed initial state from one round of measurements}
\label{sm:exp-prep}

The PDS is $|{+}\rangle^{\otimes L}$ itself, and the MMIS follows from it by one round of measurements. Measuring $Z_iZ_{i+1}$ on the bonds of the ring in $|{+}\rangle^{\otimes L}$ gives outcomes $\mu_{i,i+1}$ with $L-1$ independent values. Given $\mu$, the state is $|{\rm cat}_\mu\rangle\equiv (|s_\mu\rangle+|{-s_\mu}\rangle)/\sqrt2$ with $s_is_{i+1}=\mu_{i,i+1}$, a GHZ state in a known frame. These cat states pair the $2^L$ configurations $s$ and $-s$, so that the average over the $2^{L-1}$ equiprobable $\mu$ is exactly
\begin{equation}
2^{-(L-1)}\sum_\mu|{\rm cat}_\mu\rangle\langle{\rm cat}_\mu|=2^{-L}\bigl(\openone+\bar X\bigr),
\label{smround}
\end{equation}
the MMIS. In this protocol $\mu$ belongs to the environment, so that only the $\tau$ decoder uses the bonds of the round. This reproduces the initial-row factors of the MMIS in \cref{em:initial} and wires $\Gt$ alone. One round of measurements thus prepares the boundary condition of the MMIS derived in \cref{em:initial}. The MMIS itself is a separable mixture, consistent with \cref{em:initial}, where this boundary condition requires a mixed state rather than entanglement.

\subsection{Exact attenuation by unrecorded errors}
\label{sm:exp-err}

Errors at unknown locations are the counterpart, on a processor, of the unrecorded dephasing locations of \cref{sm:verify}, which turn the percolation problem into a disordered model. For a fixed decoder, independent stochastic Pauli faults and readout flips instead attenuate the single-shot average of \cref{smest} by an exact factor, which we derive here and evaluate at criticality. Independence is assumed at the fault level, and a single fault may affect several bonds. The independent-bond model used in the rest of this subsection is a further specialization, whose reduction from gate-level noise is tested in \cref{sm:exp-gate}.

For the fixed nonadaptive circuit, let Pauli faults $f$ after gates, idles and preparations, and flips of measured outcomes, occur independently given the event pattern, with state-independent probabilities $\zeta_f$. A Pauli fault is propagated through the subsequent gates, and it flips a measurement outcome when the propagated operator anticommutes with the measured observable. Writing each random outcome of the ideal circuit as a formal bit, every outcome is affine in these bits and in the indicators of the faults that occur, and the estimator cancels the random bits. With $[f]=1$ if the fault $f$ occurs and $0$ otherwise, as in \cref{sm:verify},
\begin{equation}
Y_\alpha=I_\alpha\,(-1)^{\sum_f\chi_f[f]},\qquad
B_\alpha=\overline{I_\alpha\prod_f\bigl(1-2\zeta_f\chi_f\bigr)},
\label{smatt}
\end{equation}
with $\chi_f\in\{0,1\}$ the \emph{sensitivity} of the path to the fault $f$. A fault that flips an even number of factors in the estimator cancels, and a channel with mutually exclusive Pauli branches contributes $\sum_Q\zeta(Q)(-1)^{\chi(Q)}$. For Bernoulli faults with $A_\alpha>0$ and $0\le\zeta_f<\frac12$ (exclusive channels need positive factors), with $\omega_f=-\ln(1-2\zeta_f)$ and the exposure $\Omega_\alpha=\sum_f\chi_f\omega_f$,
\begin{equation}
V_\alpha\equiv\frac{B_\alpha}{A_\alpha}=\bigl\langle e^{-\Omega_\alpha}\bigr\rangle_{\rm conn}
\ge e^{-\langle\Omega_\alpha\rangle_{\rm conn}},
\label{smvis}
\end{equation}
which is \cref{expvis}, with $\langle\cdot\rangle_{\rm conn}$ the average over the event patterns with $I_\alpha=1$. A bound on the mean exposure thus guarantees a window of visibility. Symmetric endpoint readout errors, independent and constant in time, only rescale $V_\alpha$. Errors of probability $\zeta_0$ on the labels of the round enter the MMIS through the $|\mathcal P_0|$ round bonds of its path, so that for fixed paths $\ln[B_\tau^{\rm MMIS}(\zeta_0)/B_\tau^{\rm MMIS}(0)]=-2\zeta_0\langle|\mathcal P_0|\rangle_b+O(\zeta_0^2)$, with the average over connected patterns weighted by $e^{-\Omega_b}$ and $\Omega_b$ the bulk part of the exposure.

For independent bond errors the path of largest visibility minimizes $\sum\omega_{ac}$ over its bonds, which Dijkstra's algorithm finds. For independent bond flips, Bayes decoding maps onto a random-bond Ising problem, as for the unrecorded locations of \cref{sm:verify}, which can improve the bias.

\begin{figure}[htb]
\centering
\includegraphics[width=5.06in]{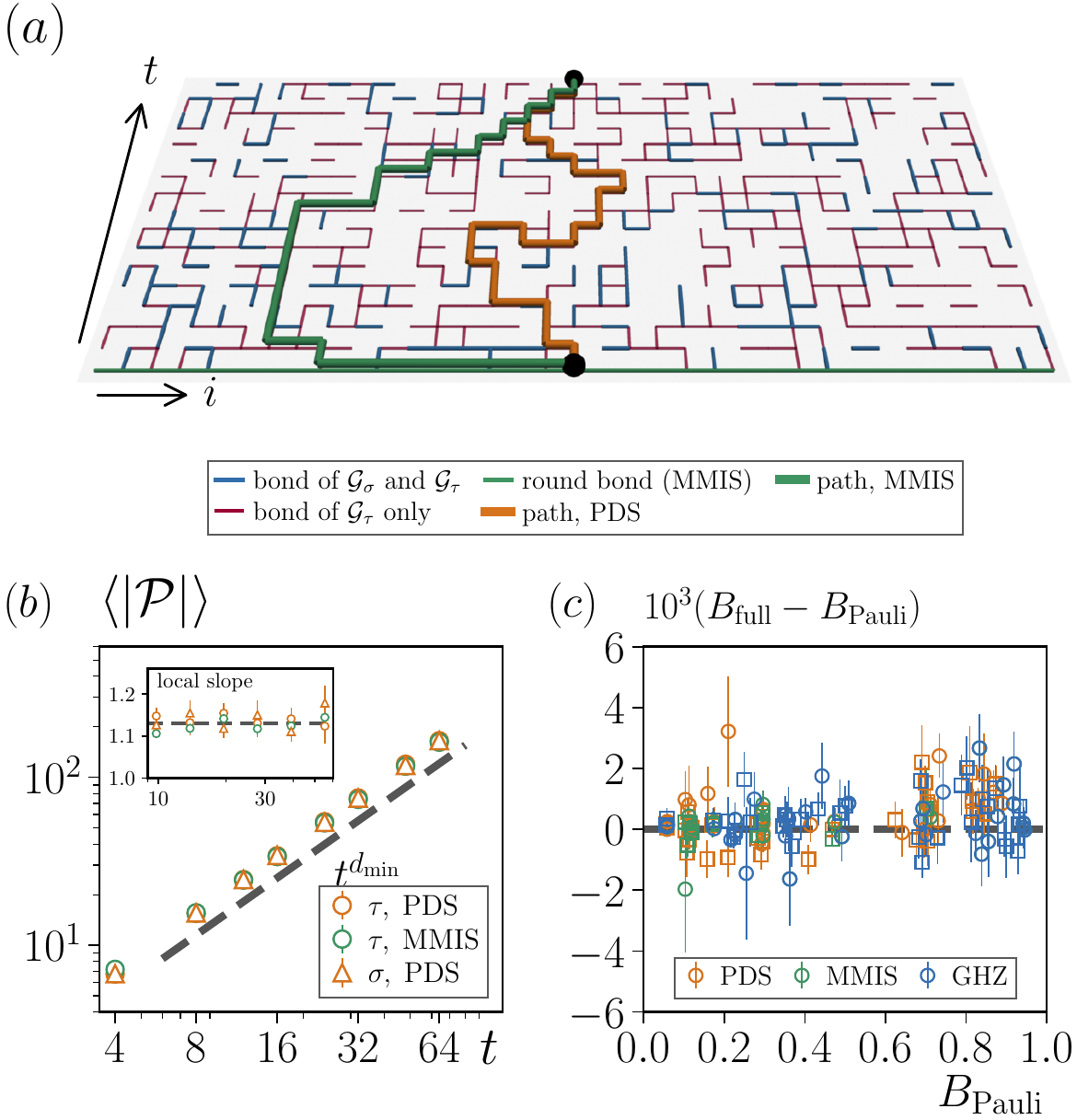}
\caption{\textbf{Decoding paths scale with the shortest-path dimension, and non-Pauli noise enters at the $10^{-3}$ level.} \textbf{(a)} One event pattern of $\Gt$ at $p^\tau_c$, sites $3$ to $37$ of $L=40$ up to $t=16$, with the minimum-weight paths from the reference $(i,0)$ to the target $(i,t)$ (black). Blue bonds belong to both graphs and thinner red bonds to $\Gt$ alone. The PDS path (orange) returns to $(i,0)$ through the bulk. The MMIS path (green) reaches the initial row elsewhere and closes along the bonds of the round. Weights are $1$ per bulk bond and $0.2$ per round bond, and the pattern was chosen for illustration. \textbf{(b)} Mean number of bulk bonds on the path over connected patterns at the critical point of each order parameter, $L=1024$, clean labels, for $\tau$ from the PDS (orange circles) and the MMIS (green circles) and for $\sigma$ from the PDS (orange triangles), with $t^{d_{\rm min}}$ dashed. Inset: slopes between successive times, with $d_{\rm min}$ dashed. \textbf{(c)} Decoded correlators under the full gate-level noise model minus the exact Pauli formula of \cref{smatt} for the matched twirl, in cases with nonzero connectivity and $L\le5$, from the PDS (orange), the MMIS (green) and the GHZ state (blue), for the ancilla (deep) and ancilla-free (pale) gadgets, with one standard error of the difference.}
\label{S3}
\end{figure}

We evaluated \cref{smatt} on rings of $L=256$ to $1024$ at both critical points, with bulk bond flips of a common probability $\zeta$, label probability $\zeta_0$, ideal endpoints, and the PDS and the MMIS paired on the same patterns. For $L=512$ and $1024$ and clean labels, the mean number $\langle|\mathcal P|\rangle$ of bulk bonds on the path grows with slopes $1.125$ to $1.142$ over $8\le t\le64$ for both order parameters and initial states, and fits with corrections give $1.08$ to $1.16$ [\cref{S3}(b)]. These values are compatible with the shortest-path dimension $d_{\rm min}=1.13077(2)$ of planar percolation~\cite{grassberger_spreading_1992,zhou_shortest-path_2012}. Each point of \cref{expt}(b,c) and \cref{S3}(b) averages over $3.3\times10^4$ event patterns at $L=1024$ and $1.3\times10^5$ at $L=512$, with one standard error. We therefore adopt the scaling hypothesis that $|\mathcal P|/t^{d_{\rm min}}$ has a limiting distribution over connected patterns for each initial state with clean labels, which gives, for $t\to\infty$ at fixed $\omega\,t^{d_{\rm min}}$,
\begin{equation}
V_\alpha\to\mathcal V_\alpha\bigl(\omega\,t^{d_{\rm min}}\bigr),\qquad t_\zeta\propto\zeta^{-1/d_{\rm min}},
\label{smcollapse}
\end{equation}
with $\mathcal V_\alpha$ a scaling function and $t_\zeta$ the time at which $V_\alpha$ falls to a fixed level. For clean labels the data for $\zeta=10^{-3}$ to $3\times10^{-2}$ collapse, and the PDS and the MMIS nearly coincide [\cref{expt}(c)]. The near coincidence reflects that the initial state selects which patterns connect but has little effect on the length of the connecting paths. At $p^\tau_c$ and $L=1024$, $V_\tau=\frac12$ is reached at $\omega\,t^{d_{\rm min}}\simeq0.47$, that is at $t\simeq16$ for $\zeta=10^{-2}$ and $t\simeq48$ for $3\times10^{-3}$, and the Jensen bound of \cref{smvis} lies within $2\%$ of $V_\tau$ in all sampled critical cases with $V_\tau>\frac12$.

For positive means the decoded contrast of \cref{expcontrast} is exactly the ideal one plus the difference of two log visibilities,
\begin{equation}
\Gamma_\tau^B\equiv\ln\frac{B_\tau^{\rm MMIS}}{B_\tau^{\rm PDS}}=\Gamma_\tau+\ln\frac{V_\tau^{\rm MMIS}}{V_\tau^{\rm PDS}},
\label{smcontrast}
\end{equation}
with $\Gamma_\tau=\ln(A_\tau^{\rm MMIS}/A_\tau^{\rm PDS})=(x_s/z)\ln t+{\rm const}$ to leading order at $p_c^\tau$ and $\kappa=0$ by \cref{laws}, in which the common observation-time dimension cancels. For the sampled clean-label cases $V_\tau^{\rm MMIS}\simeq V_\tau^{\rm PDS}$, so that the contrast is far less sensitive to bulk noise than the visibility itself. At $p^\tau_c$, $L=1024$, $\zeta=10^{-2}$ and clean labels, $\Gamma_\tau^B=1.46(2)$, $1.71(2)$ and $2.19(4)$ at $t=8$, $16$ and $64$, against $1.47(2)$, $1.71(2)$ and $2.13(3)$ without errors, a sampled bulk-noise shift of up to about $0.06$. 

Label errors act directly. Clean-label paths have $\langle|\mathcal P_0|\rangle\simeq0.49\,t$, and at $\zeta=\zeta_0=10^{-2}$ decoding with the label costs lowers the contrast by $0.07$ at $t=8$ and $0.12$ at $t=16$, less than the clean-path estimate $2\zeta_0\langle|\mathcal P_0|\rangle$, which omits the bulk weighting and the reoptimization of paths. Bulk noise thus mainly costs shots, whereas label noise shifts the signature of the initial boundary condition directly. Every path from $(i,0)$ to $(i,t)$ crosses at least $t$ temporal bonds, so that independent temporal flips of probability $0<\zeta_v<\frac12$ give $V_\alpha\le(1-2\zeta_v)^t$, which cuts off the plateau of $A_\tau$ in \cref{Sxi}(a) in the decoded correlator. For the insertion slice of \cref{Sslice}, from the PDS at $p^\tau_c$, $L=512$, $t=8$ and $\kappa=1$, the finite-window slopes of $\ln C_\tau$ against $\ln r$ over $2\le r\le8$ are $-0.50$ on the slice and $-0.85$ on the final row at $\zeta=\zeta_0=10^{-2}$, against $-0.32$ and $-0.64$ without errors. The two slopes thus remain distinct.

\subsection{Gate-level implementation and feasibility}
\label{sm:exp-gate}

We next connect gate-level noise to the bond error probabilities of \cref{sm:exp-err}. The fault-level formula \cref{smatt} is exact under its stated assumptions, whereas replacing the resulting correlated bond errors by pooled independent flips is an approximation, which we test against a full noise simulation.

We compiled both $ZZ$ gadgets of \cref{sm:exp-est} into controlled-$Z$, Hadamard and $Z$-measurement layers on an open chain. At every location of the schedule we inserted depolarization, readout assignment errors, and relaxation and dephasing, with parameters typical of current superconducting processors: two-qubit depolarizing errors of order $10^{-3}$, assignment errors of order $10^{-2}$, readouts of $1$ to $2\,\mu$s and coherence times of order $10^2\,\mu$s. Leakage, crosstalk and measurement back-action beyond assignment are not modeled. Propagating each Pauli fault through the compiled circuit gives its sensitivity to every bond, and pooling per orientation at $p^\tau_c$ gives error probabilities $\zeta_v$, $\zeta_h$ and $\zeta_0$ of about $(1\text{--}3)\times10^{-2}$ for the temporal, spatial and label bonds of either gadget.

Two checks support this reduction. On open chains of $L=32$ and $64$ up to $t=16$, the independent-bond prediction on a given path was $0.71$ to $1.00$ times the exact gate-level value, since correlated faults cancel in pairs. This ratio was observed on the tested paths and is not a bound for the large rings considered next. Statevector quantum-trajectory simulations of the scheduled circuits for $L\le5$ and $t\le4$, at most nine qubits including ancillas, with amplitude damping, dephasing and asymmetric readout, deviate from \cref{smatt} for the matched Pauli twirl by up to about $3.2\times10^{-3}$ [\cref{S3}(c)]. At these rates, sizes and times, the non-Pauli parts of the noise therefore enter at the $10^{-3}$ level.

Independent bond errors at the pooled probabilities, labels and endpoints included and paths reoptimized, give $V_\tau\simeq0.48$--$0.61$ at $t=8$ on rings of $L=1024$ at $p^\tau_c$ for both gadgets and initial states. Dividing all bond and endpoint error probabilities by four keeps $V_\tau>\frac12$ through sampled times $t\le24$. This pooled bond model thus retains visibility over several time steps. The shot cost for estimating $B_\alpha$ is given below \cref{smcost}, and at fixed relative error it exceeds the ideal one by the factor $(V_\alpha^{-2}-A_\alpha)/(1-A_\alpha)$, approximately $3$ to $7$ at $t=8$. With direct endpoint measurements, an open chain uses $L$ qubits without ancillas, or $2L-1$ with one reusable ancilla per bond. A dephasing followed by a measurement of the same $Z_iZ_{i+1}$ is a measurement, as noted below \cref{ops}, so that the expected logical controlled-$Z$ count is $2(L-1)(tp_h+1)$ for the MMIS and $2(L-1)$ fewer for the PDS. Here $p_h=17/30$ is the spatial bond probability of $\Gt$ at $p_c^\tau$ [\cref{B1}].

The required gates and mid-circuit readout have been demonstrated on several platforms. Superconducting processors offer fast gates and repeated mid-circuit readout, although relaxation and readout errors limit the observable time window~\cite{krinner_repeated_2022,google_suppressing_2023}. Trapped-ion processors provide high-fidelity gates, flexible connectivity and mid-circuit measurement, at the cost of additional transport and cooling overhead~\cite{pino_demonstration_2021,moses_race-track_2023}. Reconfigurable neutral-atom arrays offer large registers and programmable connectivity, while repeated measurements require managing atom loss and transport to readout zones~\cite{bluvstein_logical_2024}. The visibility and sampling cost on each platform follow from its own noise model through \cref{smatt}.

\end{document}